\documentclass[aps,prd,twocolumn,superscriptaddress,nofootinbib,10pt,floatfix,preprintnumbers]{revtex4-1}

\usepackage[utf8]{inputenc}
\usepackage{graphicx,amsmath,amsfonts,amssymb,aas_macros,slashed}
\usepackage{bbold}
\usepackage{datetime}
\usepackage{numprint}
\usepackage{enumitem}
\usepackage{hyperref}
\usepackage[colorinlistoftodos]{todonotes}
\usepackage{xcolor}
\usepackage{rotating}
\allowdisplaybreaks

\newbox\tablebox    \newdimen\tablewidth
\def\leaderfil{\leaders\hbox to 5pt{\hss.\hss}\hfil}
\def\endPlancktable{\tablewidth=\columnwidth 
    $$\hss\copy\tablebox\hss$$
    \vskip-\lastskip\vskip -2pt}

\def\tablenote#1 #2\par{\begingroup \parindent=0.8em
    \abovedisplayshortskip=0pt\belowdisplayshortskip=0pt
    \noindent
    $$\hss\vbox{\hsize\tablewidth \hangindent=\parindent \hangafter=1 \noindent
    \hbox to \parindent{$^#1$\hss}\strut#2\strut\par}\hss$$
    \endgroup}
\def\doubleline{\vskip 3pt\hrule \vskip 1.5pt \hrule \vskip 5pt}

\usepackage{pgfplots}
\usepackage{bm}
\usepackage{mathtools}

\makeatletter
\newcommand{\thickhline}{%
    \noalign {\ifnum 0=`}\fi \hrule height 1pt
    \futurelet \reserved@a \@xhline
}
\newcolumntype{"}{@{\hskip\tabcolsep\vrule width 1pt\hskip\tabcolsep}}
\makeatother

\begin{document}

\title{Early or phantom dark energy, self-interacting, extra, or massive neutrinos, primordial magnetic fields, or a curved universe:  An exploration of possible solutions to the $H_0$ and $\sigma_8$ problems}

\preprint{UCI-HEP-TR-2022-06}

\author{Helena\ Garc\'ia\ Escudero}
\email{garciaeh@uci.edu}
\affiliation{Department of Physics and Astronomy,  University of California, Irvine, California 92697-4575, USA}

\author{Jui-Lin\ Kuo}
\email{juilink1@uci.edu}
\affiliation{Department of Physics and Astronomy,  University of California, Irvine, California 92697-4575, USA}

\author{Ryan\ E.\ Keeley}
\email{rkeeley@ucmerced.edu}
\affiliation{Department of Physics, University of California, Merced, California 95343, USA}

\author{Kevork\ N.\ Abazajian}
\email{kevork@uci.edu}
\affiliation{Department of Physics and Astronomy,  University of California, Irvine, California 92697-4575, USA}
\begin{abstract}
    In this paper we explore the existing tensions in the local cosmological expansion rate, $H_0$, and amplitude of the clustering of large-scale structure at $8\, h^{-1}\mathrm{Mpc}$, $\sigma_8$, as well as models that claim to alleviate these tensions. We consider seven models: evolving dark energy ($w$CDM), extra radiation ($N_\mathrm{eff}$), massive neutrinos, curvature, primordial magnetic fields (PMF), self-interacting neutrino models, and early dark energy (EDE). We test these models against three datasets that span the full range of measurable cosmological epochs, have significant precision, and are well-tested against systematic effects: the Planck 2018 cosmic microwave background data, the Sloan Digital Sky Survey baryon acoustic oscillation scale measurements, and the Pantheon catalog of type Ia supernovae. We use the recent SH0ES $H_0$ measurement and several measures of $\sigma_8$ (and its related parameter $S_8=\sigma_8\sqrt{\Omega_\mathrm{m}/0.3}$). We find that four models are above the ``strong'' threshold in Bayesian model selection, $w$CDM, $N_\mathrm{eff}$, PMF, and EDE. However, only EDE also relieves the $H_0$ tension in the full datasets to below 2$\sigma$. We discuss how the $S_8/\sigma_8$ tension is reduced in recent observations. However, even when adopting a strong tension dataset, no model alleviates the $S_8/\sigma_8$ tension, nor does better than $\Lambda$CDM in the combined case of both $H_0$ and $S_8/\sigma_8$ tensions. 
\end{abstract}
\maketitle

\section{Introduction}
So far, the best-fitting scenario for describing our Universe on large scales is the standard model of cosmology, also known as $\Lambda$CDM. 
Its success in simultaneously explaining cosmological observables at low and high redshift is undeniable~\cite{Bull:2015stt}; nevertheless, in this framework several tensions in different datasets, e.g., between the cosmic microwave background (CMB) and observations at low redshift including the distance ladder and large-scale structure (LSS), have emerged. 
One of these discrepancies is the ``$H_0$ tension", which is a mismatch between the present expansion rate of the Universe, i.e., the Hubble constant $H_0$, inferred from the distance ladder built from Cepheid variables and Type Ia supernovae (SN Ia), and $H_0$ inferred from  the angular power spectra of the CMB, given a Friedmann $\Lambda$CDM cosmology evolution to today.

Recently, this conflict has grown to a level of approximately $\sim\!5 \sigma$ provided that $H_0 = 67.36 \pm 0.54\,{\rm km/s/Mpc}$ from Planck CMB data, within the $\Lambda$CDM model~\cite{Planck:2018vyg}, largely deviates from $H_0 = 73.04 \pm 1.04\,{\rm km/s/Mpc}$, reported by the SH0ES collaboration using the Cepheid-based distance ladder~\cite{Riess:2021jrx}.
Another anomaly arises when measuring $\sigma_8$, which is the value of the root-mean-square fluctuation of density perturbations calculated with a top-hat window function of $k = 8\, h^{-1}\,{\rm Mpc}$. The value $\sigma_8$ is often combined with the parameter it is most degenerate with in the combination $S_8 \equiv \sigma_8\sqrt{\Omega_\mathrm{m} /0.3}$, with $\Omega_\mathrm{m}$ being the matter density parameter.
The value of $S_8$ inferred from Planck CMB data within the $\Lambda$CDM framework, $S_8 = 0.832 \pm 0.013$~\cite{Planck:2018vyg}, and low-redshift probes such as weak gravitational lensing and galaxy
clustering~\cite{DES:2021bvc,Vikhlinin:2008ym,Planck:2013lkt,SPT:2021efh,Schellenberger:2017wdw,Heymans:2013fya,KiDS:2020suj,DES:2021bwg,Poulin:2018zxs} do not agree with the value inferred from the CMB at a statistical level from approximately $2\sigma$ to $4\sigma$~\cite{Douspis:2018xlj}.

These cosmological inconsistencies may originate from unaccounted systematic errors in the local distance ladder measurements and/or in the Planck observations. 
Extended experimental work has been carried out to determine if unknown systematics are the main reason for this mismatch. 
For instance, errors in SN Ia dust extinction modeling and intrinsic variations~\cite{Mortsell:2021nzg,Mortsell:2021tcx,Wojtak:2022bct}, Cepheid metallicity correction~\cite{Efstathiou:2020wxn} and different types of SN Ia populations are potential candidates for these systematic effects; see~\cite{DiValentino:2021izs} for a complete review. 
Additional methods of calibrating the distance ladder, such as using the J-region asymptotic giant branch~\cite{Lee:2022akw}, or calibration via gravitational-wave ``standard siren"~\cite{Chen:2017rfc} may provide an independent measure and test of the tension present in local to high-redshift determinations of $H_0$. 
In the meantime, it is of value to explore in detail the nature of new physics beyond $\Lambda$CDM that can be a robust solution to the $H_0$ tension, as well as models that aim to solve the $S_8$ tension, independently or in concert with $H_0$. 
That is what we explore here.

Depending on the cosmic period that the new physics takes effect, proposed models can be categorized into late-time and early-time solutions. 
The first category changes expansion history of the Universe at low redshift, while the latter modifies the physics of the early Universe before recombination; see~\cite{Schoneberg:2021qvd} for a recent review.
Late-time solutions include, for example, $w$CDM~\cite{Planck:2018vyg}, $w_0 w_a$CDM~\cite{Planck:2018vyg} or an interacting dark energy model~\cite{Lucca:2020zjb,Gomez-Valent:2020mqn}. However, given tight constraints on cosmic expansion history at low redshift, late-time solutions are in general highly disfavored as solutions to $H_0$ tension~\cite{DiValentino:2020naf,Keeley:2022ojz}.
On the other hand, early-time solutions, e.g., early dark energy (EDE)~\cite{Poulin:2018dzj,Poulin:2018cxd,Smith:2019ihp}, a modified neutrino sector~\cite{Bialynicka-Birula:1964ddi,Raffelt:1987ah,Chacko:2003dt,Atrio-Barandela:1996suw,Bell:2005dr,Sawyer:2006ju,Friedland:2007vv,Basboll:2008fx,Jeong:2013eza,Oldengott:2017fhy,Kreisch:2019yzn,Park:2019ibn,Blinov:2020hmc,He:2020zns,Das:2020xke,Esteban:2021ozz,Venzor:2022hql,Berryman:2022hds,Abazajian:2017tcc}, baryon inhomogeneity sourced from primordial magnetic fields~\cite{Jedamzik:2018itu} and extra dark radiation before recombination (e.g., Ref.~\cite{Vagnozzi:2019ezj}), are seen as better candidates in alleviating the tension by keeping $\Lambda$CDM's successes in the late Universe intact. We also consider nonzero neutrino mass as the solution to the $S_8$ tension~\cite{Moskowitz_2014,Battye:2013xqa,Wyman:2013lza,BOSS:2014etx,Mccarthy:2017yqf}, both on its own and in tandem with other new physics related to both tensions.

Based on established statistical methods for model rejection, we explore a collection of new physics models proposed to alleviate the tensions. Many existing and new theoretical proposals in the literature only judge a new model relative to standard $\Lambda$CDM, and sometimes by only comparing the inferred central values of $H_0$ or $S_8$ between $\Lambda$CDM and the new model. 
Furthermore, the effects of new models on several other robust cosmological datasets go unaddressed, including the baryon acoustic oscillation (BAO) feature and detailed accelerated expansion history at low redshifts, measured by SN Ia. 
Meanwhile, new results in observational cosmology often explore only one or two example excursions from $\Lambda$CDM. In our work we combine a large set of proposed tension-reduction models with the latest robust observational cosmological data in order to assess which models may successfully resolve the tension while being consistent with the available hallmark data. 
Along these lines, we consider  and evaluate, in detail, the specific statistical significance of any remaining $H_0$ and $S_8$ tensions in proposed models, separately and in concert.  
In summary, the objective of this work is finding the best model, or models, proposed so far that agree with measurements that indicate these anomalies.

This paper is organized as follows: in Sec.~\ref{sec:model}, we list the beyond $\Lambda$CDM models studied in this work and discuss the way that they reduce cosmic tensions. In Sec.~\ref{sec:methodology}, we give details of and motivations for the datasets included in our calculations and the statistical strategies and computational tools employed for deriving statistical significance.
We present and discuss the results of the tests made in Sec.~\ref{sec:result}. In Sec.~\ref{sec:discussion}, we analyze the results and discuss the physics of cosmological parameters' shift for different models and tension datasets, relative to the $\Lambda$CDM fit to the CMB. Finally, we summarize the main conclusions of this work in Sec.~\ref{sec:conclusion}.

\section{Beyond $\Lambda$CDM models}
\label{sec:model}

In this section, we discuss the beyond $\Lambda$CDM models considered in this work.
We briefly sketch the physics of each model that alleviate the cosmic tensions and we refer readers to Appendix~\ref{app:model} for details of the models.

\begin{itemize}
\item $w$CDM: In the dark energy domination era, phantom dark energy with equation of state $w < -1$ can further accelerate the expansion of the Universe compared to the standard $w= -1$ case. Therefore, the $H_0$ inferred from  CMB experiments can be reconciled with the $H_0$ measured from local measurements~\cite{DiValentino:2016hlg,Ludwick:2017tox,Vagnozzi:2019ezj,Alestas:2020mvb}; see also~\cite{DiValentino:2017zyq,Yang:2018qmz} for different parametrizations of $w$. The evolution of dark energy density via a nonstandard equation of state $w$ also alters the growth of structure~\cite{Abazajian:2002ck,Joudaki:2017zhq,Keeley:2019esp}, which can alleviate or exacerbate the $S_8$ tension.
\item nontrivial neutrino mass, $\Sigma m_\nu > 0.06\,\mathrm{eV}$: a lower amplitude of clustering at smaller scales, and therefore smaller $\sigma_8$ or $S_8$, can be achieved by increasing the neutrino mass and its contribution to the total matter density~\cite{Hu:1997mj}. Several papers have suggested an indication of nonzero active neutrino masses, or combinations of extra mass eigenstates and neutrino masses because of low-$\sigma_8$ measurements, e.g., Refs.~\cite{Joudaki:2012uk,Battye:2013xqa,Wyman:2013lza,Dvorkin:2014lea,BOSS:2014etx}.
\item $\Lambda$CDM+$N_{\rm eff}$: A correlation exists between the Hubble parameter inferred from CMB measurements and the radiation energy budget in the early Universe. The latter can be parametrized by the effective number of relativistic degrees of freedom $N_{\rm eff}$. Therefore, extra relativistic species beyond the standard model neutrinos, such as dark radiation, can effectively reduce the sound horizon, i.e., increase $H_0$. Additional relativistic energy density affects the position of the acoustic peaks of the CMB relative to the photon damping scale, both of which are well constrained by measurements of the CMB. Extra (sterile) neutrino mass eigenstates can mimic relativistic degrees of freedom at early times and contribute to $\Sigma m_\nu$ at late times, and therefore may combine the effects of $\Sigma m_\nu$ and $N_{\rm eff}$. See, e.g., a review in Ref.~\cite{Abazajian:2017tcc}. 
\item Nonzero curvature: The size of angular diameter distance, which is measured in low-redshift measurements such as BAO and SNe, is closely related to the curvature of Universe. Therefore, allowing a nonflat Universe, i.e., making the density parameter of curvature $\Omega_k$ a free parameter, offers an additional degree of freedom to modify the low-redshift spacetime geometry. Nonzero curvature also alters the growth of structure, potentially alleviating the $S_8$ tension. A nonzero curvature can be integrated into models that modify the early Universe to better fit low-redshift measurements; see Refs.~\cite{Ryan:2018aif,Handley:2019tkm,DiValentino:2019qzk} for example. Models that modify the electron mass~\cite{Sekiguchi:2020teg} along with added curvature are highly constrained by primordial nucleosynthesis~\cite{Seto:2022xgx}, so we do not consider them here.

\item Early dark energy: A potential solution of $H_0$-tension is early dark energy (EDE)~\cite{Poulin:2018dzj,Poulin:2018cxd,Smith:2019ihp}, which behaves like a cosmological constant with an equation of state $-1$ and makes up a non-negligible fraction of the energy budget before a critical redshift $z_c$.
At $z < z_c$, the energy density of EDE dilutes faster than radiation. %
By requiring $z_c$ being larger than the redshift of recombination, the expansion rate is boosted at $z > z_c$ while leaving the cosmology at $z < z_c$ intact. 
Therefore, the sound horizon is reduced such that the inferred value of $H_0$ is larger which reconciles the result from early- and late-time observations. In our work, we adopt the EDE model of Smith \textit{et al.}~\cite{Smith:2019ihp} as it can provide a better fit to the high-$\ell$ $C_\ell$ of Planck 2018.
\item Self-interacting neutrinos (SI$\nu$): In the standard cosmology, it is well known that neutrinos free-stream after the decoupling from the Standard Model (SM) thermal bath, damping the perturbations below the corresponding free-streaming scale.
It has been proposed that increasing the relativistic degrees of freedom and/or introducing a nonzero neutrino mass can help in alleviating the Hubble tension; however, these kinds of scenarios also result in a stronger suppression on perturbations due to the free-streaming of neutrinos and relativistic particles. 
To counteract the damping effect, one can consider including nonstandard interactions of the relativistic species, which delay the self-decoupling and the ensuing free-streaming~\cite{Bialynicka-Birula:1964ddi,Raffelt:1987ah,Chacko:2003dt,Atrio-Barandela:1996suw,Bell:2005dr,Sawyer:2006ju,Friedland:2007vv,Basboll:2008fx,Jeong:2013eza,Oldengott:2017fhy,Kreisch:2019yzn,Park:2019ibn,Blinov:2020hmc,He:2020zns,Das:2020xke,Esteban:2021ozz,Venzor:2022hql,Berryman:2022hds}.
In this model, alleviation comes from self-interaction of the neutrinos plus extra relativistic neutrinos that are introduced by the self-interacting mechanism itself, e.g., with seclusion of the mediating particle, its becoming nonrelativistic, and its recoupling by transfer of its energy density to the neutrinos~\cite{Chacko:2003dt}. Specifically, the moderate interaction level has been shown to be preferred by the data~\cite{Oldengott:2017fhy,Kreisch:2019yzn,Park:2019ibn,Das:2020xke}, which we confirmed in our analysis.  Therefore, our baseline model for SI$\nu$ is enhanced neutrino self-interactions at the moderate level, plus $N_\mathrm{eff}$.
\item Primordial magnetic fields (PMF) \& baryon inhomogeneity: The existence of primordial magnetic fields can introduce baryon inhomogeneities in the early Universe, which enhances the hydrogen recombination rate compared to the standard scenario~\cite{Jedamzik:2018itu}. As a result, CMB photon decoupling happens earlier and the sound horizon is reduced. Assuming the late-time evolution of the Universe is unchanged, the inferred value of $H_0$ from CMB becomes closer to that of late-universe measurements~\cite{Jedamzik:2020krr}.

\end{itemize}

\section{Methodology}
\label{sec:methodology}

\subsection{Datasets}
\label{sec:dataset}

Here, we briefly describe the cosmological datasets included in this work, and our motivation for their inclusion.  The first three observational datasets compose our baseline case for testing new physics. We choose these three as they are robust and broad: first, they are large datasets that have small to minimum-possible statistical errors; second, they have been tested extensively for systematic errors, as summarized below; and, third, are measures of cosmological parameters across the broadest possible range of cosmological history, from the last scattering surface to low-redshift:
\begin{itemize}
\item Planck 2018 CMB data (P18): for all of the calculations in this work, we use the CMB temperature and polarization angular power spectra \textit{TT,TE,EE+lowl+lowE} from the Planck 2018 legacy final release release~\cite{Planck:2018vyg}. The tension between the Planck mission's measurement of the amount of lensing existing in the temperature power spectra data have been widely studied in the last years~\cite{Planck:2013pxb,Planck:2016tof,Motloch:2018pjy}. In order to isolate the effect of low-redshift clustering measurements and their corresponding potential tension, we decided not to include the Planck CMB lensing measurements in our analysis.

\item BAO DR16 (BAO16): we include BAO data from the Sloan Digital Sky Survey (SDSS) lineage of experiments in the large-scale structure, composed of data from SDSS, SDSS-II, BOSS, and eBOSS~\cite{eBOSS:2020yzd} (combining data from BOSS DR12~\cite{BOSS:2016wmc} and eBOSS DR16). These cosmological measurements of the positions and redshifts
of galaxies provide their correlation function, which gives a tight constraint on the product of the sound-horizon scale and $H_0$. The sample consists of galaxies, quasars and Lyman-$\alpha$ forest samples' measurement of the BAO sound-horizon scale, making this combination the largest and most constraining of its kind. We included the first 2 redshift bins of the BOSS DR12 luminous red galaxy (LRG) likelihoods in the redshift range $0.2<z<0.6$, as well as the eBOSS DR16 LRG, quasar, Lyman-$\alpha$ forest, and Lyman-$\alpha$ forest-quasar cross correlation likelihoods in the redshift range $0.6<z<2.2$. These BAO datasets have been extensively tested with mock catalogs in their determination of the correlation function measurement of the BAO scale with respect to systematic theoretical  uncertainties, including fiducial cosmology, satellite galaxy kinematics, dynamics,  associated redshift space distortions, and methodological uncertainties, including clustering estimators, random catalogues, fitting templates, and covariance matrices~\cite{Alam:2020jvh,Avila:2020rmp,Vargas-Magana:2016imr,Rossi:2020wxx,Smith:2020stf}.
\item Pantheon Sample (SN): we include the Pantheon 2018 SN Ia sample from Ref.~\cite{Pan-STARRS1:2017jku}, which combines SDSS, SNLS, and low-redshift and Hubble Space Telescope samples to form the largest sample of SN Ia. In total, the sample consists of 1048 SN Ia in the redshift interval $0.01 \leq z \leq 2.3$. Moreover, this sample includes improvements, such as corrections for expected biases in light-curve fit parameters and their errors, which have substantially reduced the systematic uncertainties related to photometric calibration.
\end{itemize}
Next, we use the latest SH0ES measurement of the local Hubble constant:
\begin{itemize}
    \item SHOES H0 measurement (R21): we include a Gaussian likelihood of the Hubble constant inferred by the measurements obtained by the SH0ES collaboration in~\cite{Riess:2021jrx}, $H_0 = 73.04 \pm 1.04\,{\rm km/s/Mpc}$. 
\end{itemize}
For the possible tension with clustering on small scales, we explore a range of cluster and lensing data:

\begin{itemize}
    \item X-ray Clusters (V09): we include constraints on the cosmological parameters from the Viklinin (2009) \cite{Vikhlinin:2008ym} measurement on the galaxy cluster mass function in the redshift interval $z = [0,0.9]$. The observations of 86 x-ray clusters led to a determination of stringent constraints on the cosmological parameters, thanks to the higher statistical accuracy and smaller systematic errors than these datasets had ever reached. We use the constraints presented in Table I in Ref.~\cite{Vikhlinin:2008ym}, $\Omega_\mathrm{m} h = 0.184 \pm 0.024\,$,  $\sigma_8 (\Omega_\mathrm{m} / 0.25)^{0.47} = 0.813\pm 0.013$ and $\Omega_\mathrm{m} = 0.34 \pm 0.08$. This dataset constraint is a high-precision determination of these parameters, and sets into place one of the biggest tensions on the cosmological parameter $\sigma_8$. Due to its high precision, tension, and use in previous studies to indicate new physics, as well as our desire test if any candidate model can alleviate both inconsistencies simultaneously, we include this dataset as a key determinant of the $S_8$ problem.

     \item SZ Clusters (SZ21): we include results from the 2021 release of the SPT-SZ survey which show that within  $\Lambda$CDM, the SPT-SZ cluster sample  prefers $\sigma_8 \sqrt{\Omega_\mathrm{m} /0.3}  = 0.794\pm 0.049$, without the Planck power spectrum measurement considered in Ref.~\cite{SPT:2021efh}, as we consider P18 separately. The sample of 513 clusters from an SPT SZ sample combined with other analyzed x-ray and weak lensing samples have made this catalog one of the largest, with several methods of determining the cluster observable-mass relation.
     
    \item Dark Energy Survey Year 3 results (DES): we include results from the most recent DES Y3 survey. The photometric redshift calibration methodology they use is the first of its kind, able to recover the true cosmology in simulated surveys, encompassing information from photometry, spectroscopy, clustering cross-correlations and galaxy–galaxy lensing ratios. It employed a combination of 18 synthetic galaxy catalogs designed for the validation of combined clustering and lensing analyses. We use the cosmological constraints   $S_8  = 0.813 ^{+ 0.023} _{ -0.025}$ and $\Omega_\mathrm{m} = 0.290 ^{+ 0.039} _{ -0.063}$ obtained from their analysis~\cite{DES:2021bwg}.

\end{itemize}

We note that not all of these datasets are used for all of the statistical tests and cosmological parameters space analyses demonstrated in Sec.~\ref{sec:result}.
Therefore, to avoid confusion we will denote the datasets used for each figure and for more extensive model studies presented later.
\subsection{Statistical and cosmological software}

In our analysis, we use two different statistical tests in order, first, to quantify the success of each $\Lambda$CDM extension, and second, to measure the tension with respect to the $S_8$ and $H_0$ measurements. The two aforementioned strategies are explained in the following.
For the datasets P18, P18+BAO16 and P18+BAO16+SN, alone and also adding the $H_0$ and $S_8$ (V09, DES, SZ21) constraints, we compute the change in the effective minimal chi-square $\chi^2_{\rm min} = - 2 \ln{\mathcal{L}} $ where $\mathcal{L}$ represents the maximum likelihood for the considered model $\mathcal{M}$.
The $\Delta \chi^2_{\mathcal{M}}$ relative to $\Lambda$CDM is then derived as
\begin{align}
    \Delta \chi^2_{\mathcal{M}} \equiv \chi ^2_{{\rm min},\mathcal{M}} -\chi^2_{{\rm min},\Lambda{\rm CDM}}\,.
\end{align}
The $\chi^2$ value of a dataset can be used determine if a trend in the data is happening due to chance or due to a new model component, and can also be used to test a model's ``goodness of fit''~\cite{10.2307/1402731}. 
However, the $\Delta \chi^2_{\mathcal{M}}$ test does not take into account the complexity of each model, i.e., number of parameters it has.
Thus, we also adopt the Akaike information criterion (AIC) that allows fair comparison between models with a different number of parameters. 
In order to assess the extent to which the fit is improved, for each model we compute the AIC value~\cite{1100705} defined as ${\rm AIC} = -2 \ln{\mathcal{L}} +2k$,  with $k$ being the number of parameters of the model. 
A model is more preferred, relative to a different model, if it decreases the AIC. 
To compare with $\Lambda$CDM, we calculate the AIC of $\mathcal{M}$ relative to that of $\Lambda$CDM, defined as
\begin{align}
    \Delta {\rm AIC} \equiv \Delta \chi^2_\mathcal{M} + 2(N_{\mathcal{M}} - N_{\Lambda{\rm CDM}} )\,,
\end{align}
where $N_{\mathcal{M}}$ and $N_{\Lambda{\rm CDM}}$ represent the number of free parameters of $\mathcal{M}$ and $\Lambda{\rm CDM}$, respectively.
It is worth highlighting that this method penalizes models which introduce new parameters that do not improve the fit; therefore, a model with a lower AIC value is more successful theoretically and statistically than one with a higher AIC value. 
To judge the success of each model, we interpret our AIC values against the Jeffreys' scale~\cite{jeffreys1998theory}. This is an empirically calibrated scale with variation in adjectival description of the evidence limits. We choose a categorically ``strong" threshold of $p^{-1} = 10^{3/2}$, or 30:1 odds. This is the same criteria used in other recent works found in the literature, e.g.,~\cite{Schoneberg:2021qvd}. Our choice for a preferred model  $\mathcal{M}$  over $\Lambda$CDM places it ``strong" on the Jeffreys' scale, i.e., $\Delta$AIC $<-6.91$, with a more negative AIC being a more successful model. 

Moreover, we want to quantify the tension when adding $H_0$ or $S_8$ measurements to our datasets. We calculate
\begin{align}
\label{eq:3}
    \sqrt{ \Delta \chi^2_\mathcal{D}} \equiv  \sqrt{\chi^2_{{\rm min},\mathcal{D+T}} -\chi^2_{{\rm min},\mathcal{D}}} \,,
\end{align} 
where the subindex ${\mathcal{D}}$ represents the baseline datasets considered and  ${\mathcal{T}}$ represents the tension constraints added in the minimization calculation. The value of $\chi^2_{{\rm min},\mathcal{T}}$ is zero in our case. It is important to highlight that this particular test does not compare the goodness of a particular model in describing the data, it just quantifies the tension of a certain model when adding additional data. Therefore, using this strategy one can measure the tension level in units of standard deviation, $\sigma$. 

We explore the posterior distributions of cosmological and derived parameters of the preferred models using the publicly available Bayesian analysis framework \verb|cobaya|~\cite{Torrado:2020dgo}, with the Markov chain Monte Carlo (MCMC) sampler~\cite{Lewis:2002ah,Lewis:2013hha} and fastdragging~\cite{2005math......2099N}. For the best-fit likelihood and parameter calculation we use the minimizer sampler available in \verb|cobaya|~\cite{2018arXiv180400154C,2018arXiv181211343C,Powell_2009}. We choose flat priors with cutoffs well outside where likelihoods are significant.

\section{Results}
\label{sec:result}

\begin{figure}[t]
    \centering
    \includegraphics[width=\columnwidth]{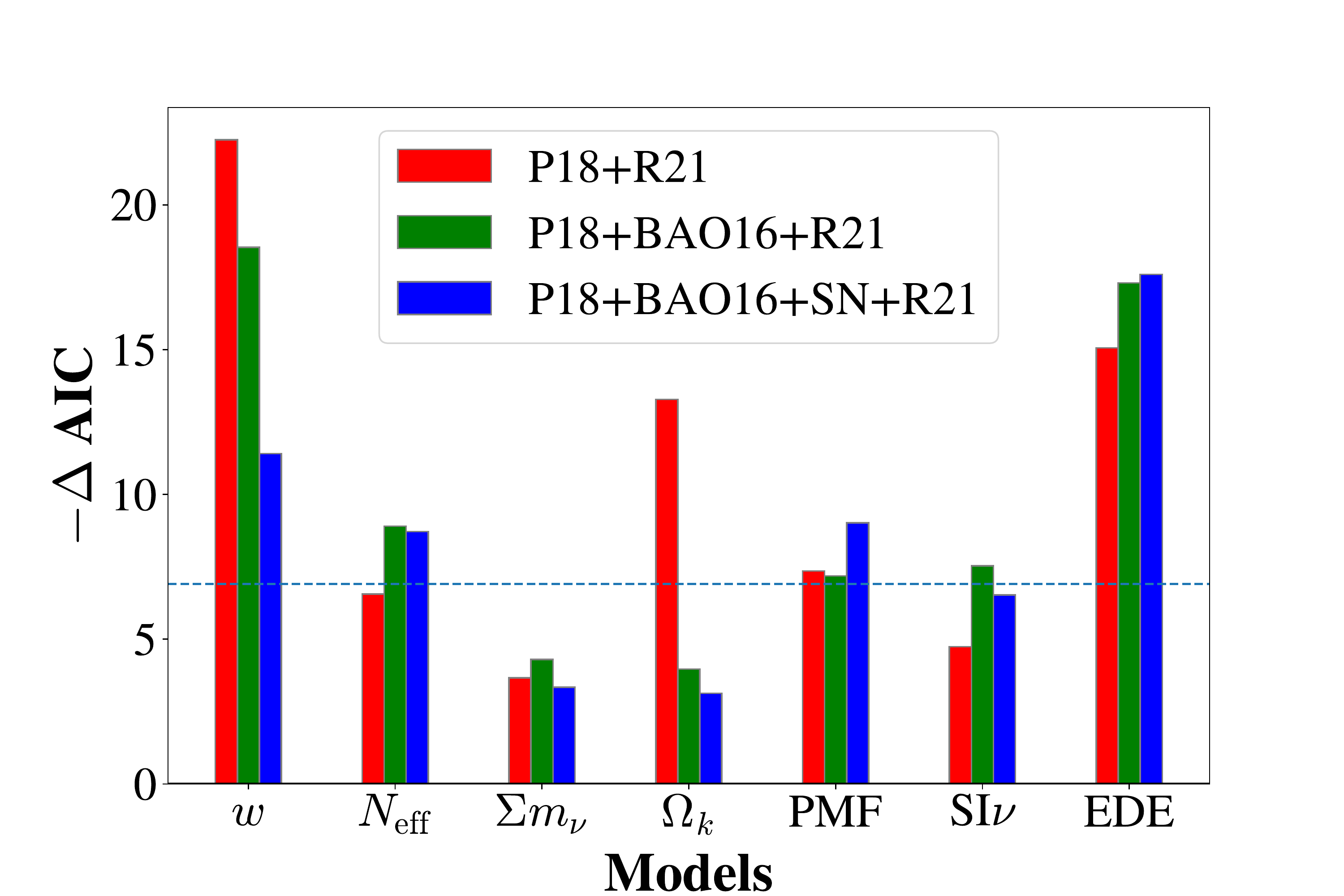}
    \caption{Shown here are the $-\Delta$AIC values of considered models (from left to right: $w$CDM,  $\Lambda$CDM+$N_{\rm eff}$, nontrivial neutrino mass, nonzero curvature, primordial magnetic fields, self-interacting neutrinos   and early dark energy) for relieving the $H_0$ tension for our baseline CMB, BAO and SNe datasets: P18+R21 (red), P18+BAO16+R21 (green) and P18+BAO16+SN+R21 (blue). The horizontal line is the threshold of stronger than ``weak preference" on the Jeffreys' scale.} 
    \label{fig:AIC_h0}
\end{figure}

\subsection{Alleviation of the $H_0$ tension}
\label{sec:H0_tension}

\begin{figure}[t]
\centering
\includegraphics[width=\columnwidth]{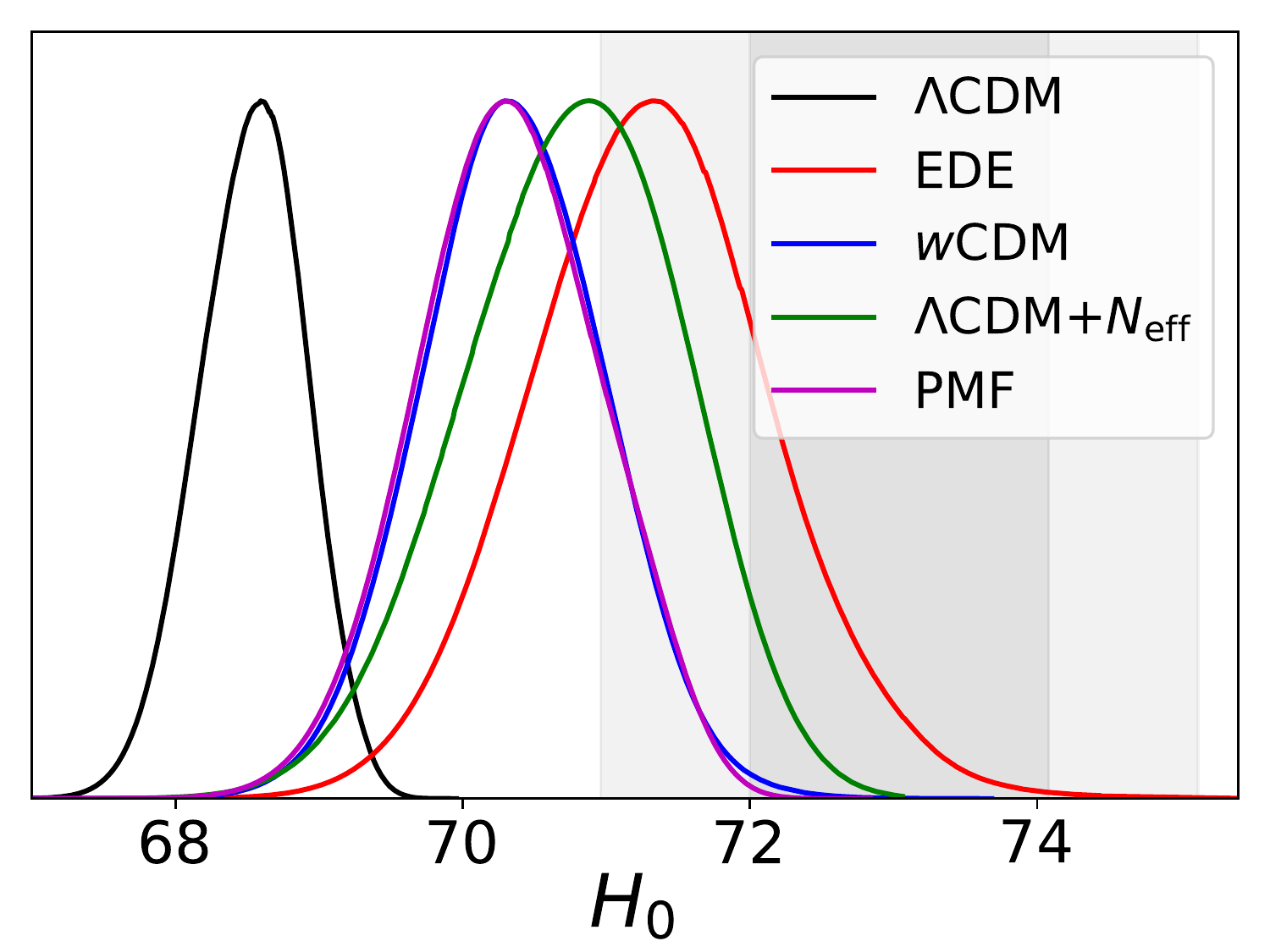}
\caption{Comparison of $H_0$ posterior distributions from early dark energy (red), $w$CDM (blue), $\Lambda$CDM+$N_{\rm eff}$ (green) and primordial magnetic fields (purple) to $\Lambda$CDM (black) being tested against P18+BAO16+SN+R21. The gray shaded regions are 68\% and 95\% C.L. limits on $H_0$ from R21.}
\label{fig:h0_comparison}
\end{figure}

The results from the AIC test, including the R21 Gaussian prior for the Hubble constant~\cite{Riess:2021jrx}, are presented in Fig.~\ref{fig:AIC_h0}. The AIC values of all candidate models are negative, which means all models do fit the considered datasets better. 
Nevertheless, when considering the full datasets P18+BAO16+SN (blue), there are only four models which cross the ``strong" threshold of  -6.91 on the Jeffreys' scale.
The most preferred model is EDE with $\Delta{\rm AIC} = -17.6$ with respect to $\Lambda$CDM (or roughly 6600:1 odds). Evolving dark energy, $w$CDM, comes in next, followed by PMF, and then CDM+$N_{\rm eff}$. The use of the supernova absolute magnitude prior ($M_b$) instead of the $H_0$ for models that behave very differently from $\Lambda$CDM at or very near redshift zero has been recently discussed in the literature~\cite{Efstathiou:2021ocp,Camarena_2021}. The only model that could have been affected by this bias from the ones we analyzed in our work is $w$CDM. We tested this model using both priors and found that the our conclusions are not affected by the use of either of them. The overall fit of both $\Lambda$CDM and $w$CDM to SH0ES is poorer when using an $M_b$ prior, but  AIC preference for $w$CDM is enhanced by 0.5 with the $M_b$ prior. Therefore, from this analysis we select EDE,  $w$CDM, PMF and  CDM+$N_{\rm eff}$ as the most successful candidates with respect to $\Lambda$CDM.
Nearly the same hierarchy is obtained when SN is not included in our calculations (green), with $w$CDM having a slight preference over EDE. 
When only P18 is taken into account (red), the preferred models change, $w$CDM becomes the preferred with $\Delta{\rm AIC} = -22.2$, followed by EDE and a nonzero curvature, significantly surpassing the reference strong threshold regime, with this $\Delta{\rm AIC}$ corresponding to roughly 66,000:1 odds. No other models cross the ``strong" threshold in this case. 
The first column of Table~\ref{tab:sigmatension} shows the number of $\sigma$ value of the residual tension of each model given our full baseline dataset P18, BAO16, and SN, when including the SHOES collaboration (R21) $H_0$ measurement. The residual is calculated using Eq.~\eqref{eq:3}. 
In addition, in Fig.~\ref{fig:h0_comparison} we show the 1D posterior distributions of $H_0$ in our preferred models relative to the R21 measurement. 

\begin{figure}[t]
    \centering
    \includegraphics[width=\columnwidth]{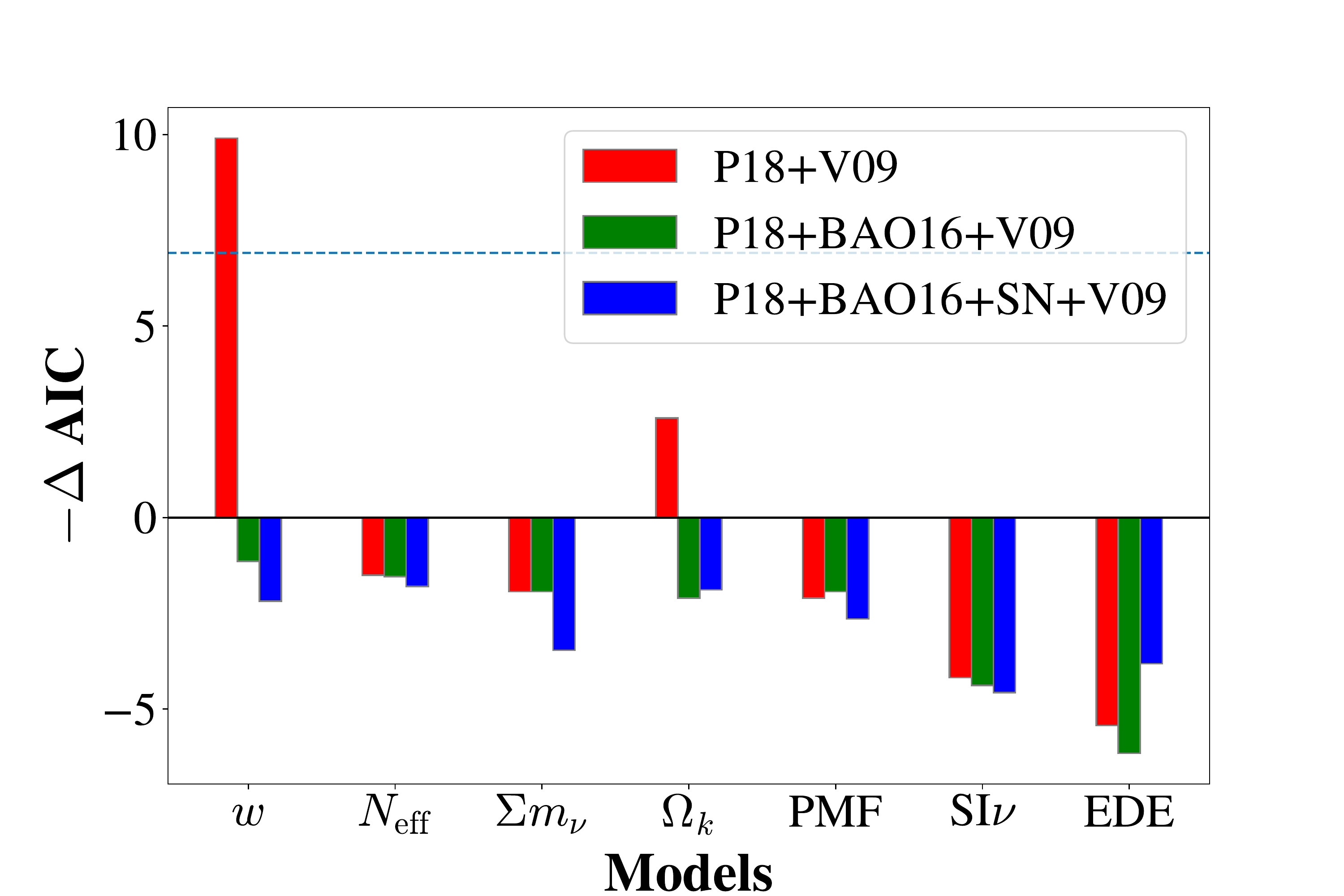}
    \caption{Shown here are the $-\Delta$AIC values of considered models (from left to right: $w$CDM,  $\Lambda$CDM+$N_{\rm eff}$, nontrivial neutrino mass, nonzero curvature, primordial magnetic fields, self-interacting neutrinos   and early dark energy) for relieving the $S_8$ tension for our baseline CMB, BAO and SNe datasets: P18+V09 (red), P18+BAO16+V09 (green) and P18+BAO16+SN+V09 (blue). The horizontal line is the ``strong" threshold on the Jeffreys' scale.}
    \label{fig:AIC_s8}
\end{figure}

\subsection{Alleviation of $S_8$ tension}
\label{sec:S8_tension}

The results of our AIC tests for the $S_8$ tension are presented in Fig.~\ref{fig:AIC_s8}. In order to force the strongest test for new physics, we choose the  most constraining and highest-tension dataset, the V09 measurements, on their combination of the $\Omega_\mathrm{m}$ and $\sigma_8$ parameters. None of the models have a negative AIC value when BAO16 alone or BAO16+SN datasets are included in the minimization. When reducing to only P18 and V09, uniquely two models are preferred: $w$CDM and a nonzero curvature, being  $w$CDM the only one crossing over the ``strong'' threshold in this case.

\begin{figure}[t]
\centering
\includegraphics[width=\columnwidth]{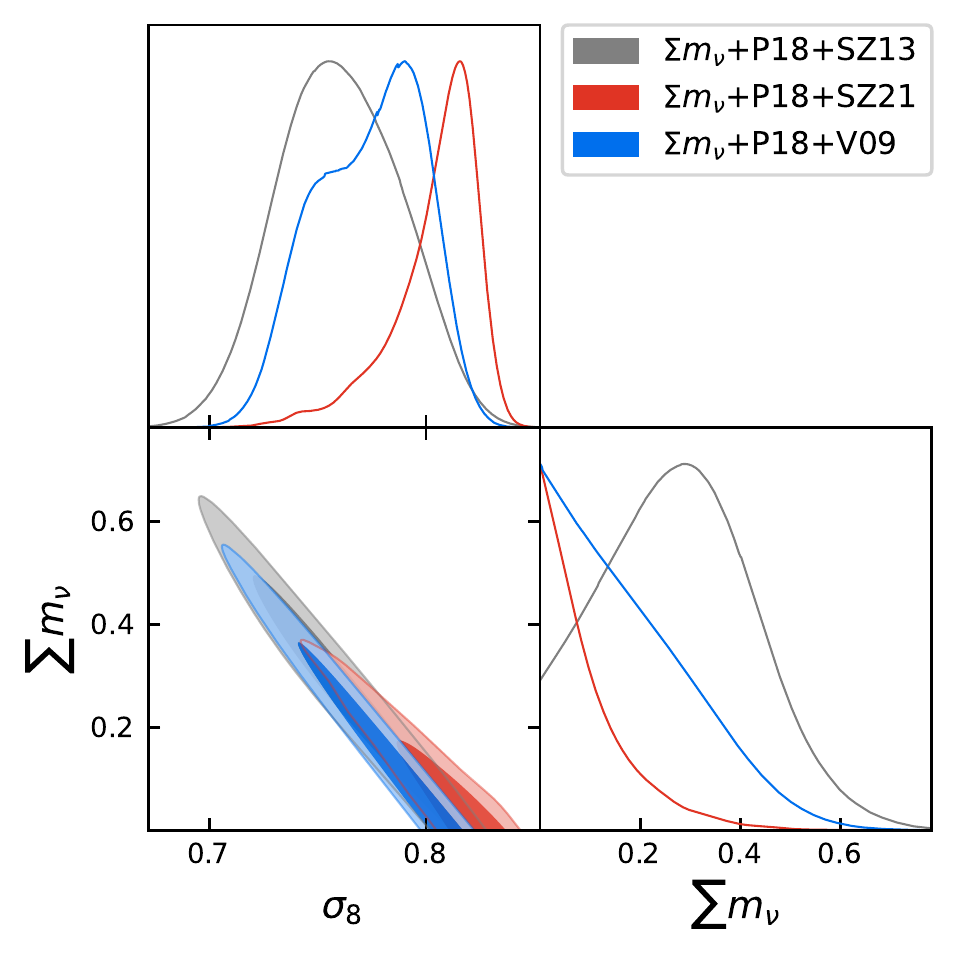}
\caption{We show 68\% and 95\% C.L. contours of ($\Sigma m_\nu$, $\sigma_8$) and their posterior distribution for the combination of P18 with our three cluster datasets: SZ13 (gray), SZ21 (red) and V09 (blue). Only P18+SZ13 has a likelihood that peaks at nonzero neutrino mass, with $\Sigma m_\nu = 0.28\pm 0.14\,{\rm eV}$.}
\label{fig:neutrino_mass}
\end{figure}

There has been significant interest in the $S_8$ tension indicating a preference for non-minimal neutrino masses (i.e., not hierarchical, but degenerate neutrino masses). Therefore, in Fig.~\ref{fig:neutrino_mass}, we show contours of $\sigma_8$ v.s. $\Sigma m_{\nu}$, comparing V09 (blue), SZ13 (gray) and SZ21 (red).
The first two datasets are in the strongest tension with P18 in terms of $\sigma_8$, while SZ21 is more consistent with P18. 
Importantly, even with the strongest tension combination, P18+V09, which prefers a slightly lower $\sigma_8\sim 0.78$, there is no preference for a nonzero neutrino mass. 
On the other hand, P18+SZ13 gives a much lower $\sigma_8 \sim 0.75$, and the likelihood peaks at a nonzero neutrino mass. The value of the neutrino mass is inferred in this dataset combination to be $\Sigma m_\nu = 0.28\pm 0.14\,{\rm eV}$, and still consistent with a minimal sum of neutrino masses of $\Sigma m_\nu = 0.06\,\mathrm{eV}$ at 2$\sigma$. 

We also ran model minimizations for SZ21, DES, and SZ13 with our baseline datasets (P18+BAO16+SN) and their subsets. None of those tension datasets had greater preference for any of the models than V09, as given by their $\Delta$AIC.
Following Eq.~\eqref{eq:3} again, we quantify the existing number of $\sigma$ tension from the $S_8$ measurement alone of V09 and DES for each model. These results are presented in the second and third columns of Table \ref{tab:sigmatension}.

\begin{figure}[t]
    \centering
    \includegraphics[width=\columnwidth]{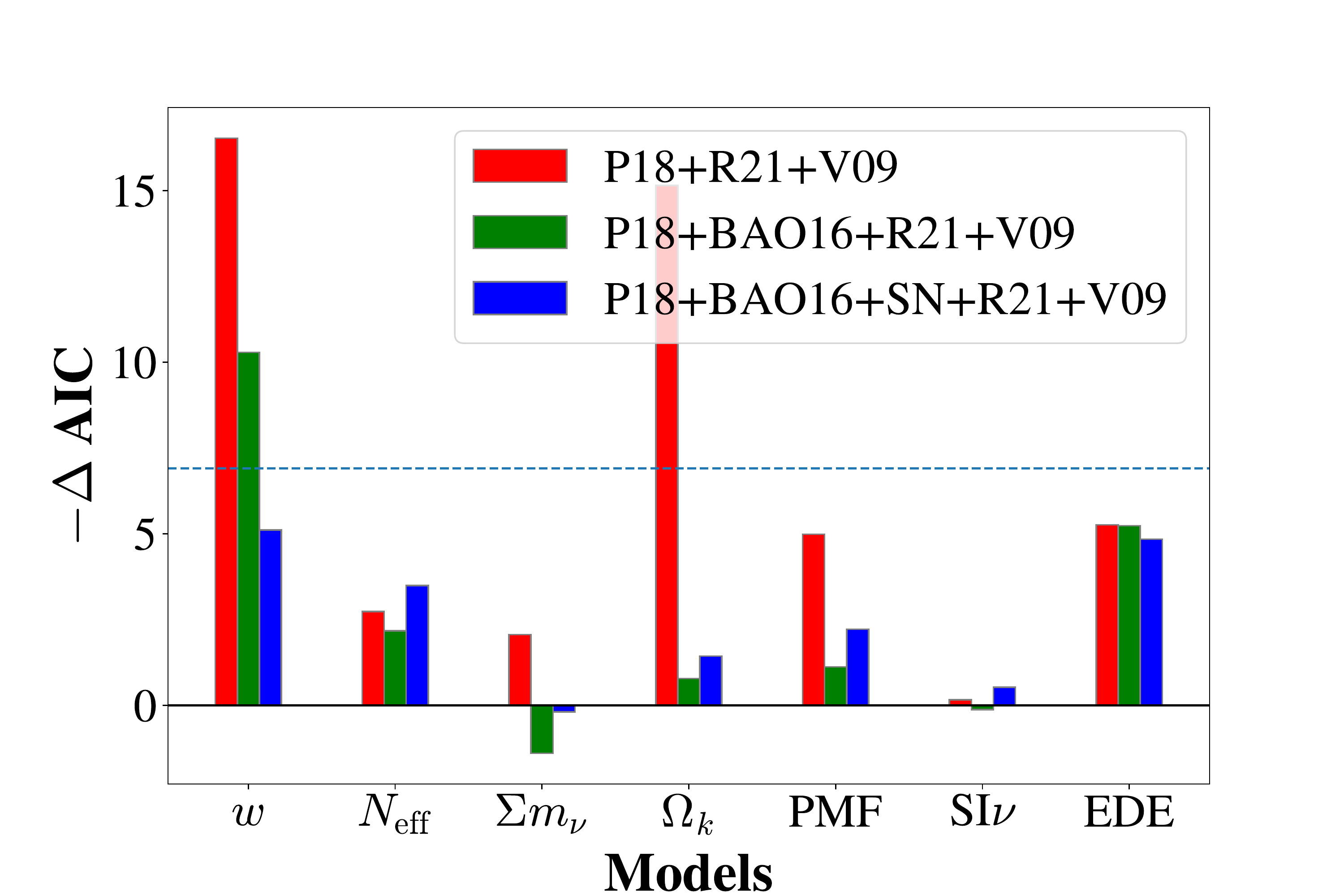}
    \caption{Shown here are the $-\Delta$AIC values of considered models (from left to right: $w$CDM,  $\Lambda$CDM+$N_{\rm eff}$, nontrivial neutrino mass, nonzero curvature, primordial magnetic fields, self-interacting neutrinos   and early dark energy) for relieving the $H_0$ and $S_8$ tensions simultaneously for our baseline CMB, BAO and SNe datasets: P18+R21+V09 (red), P18+BAO16+R21+V09 (green) and P18+BAO16+SN+R21+V09 (blue). The horizontal line crosses the threshold of  ``strong" on the Jeffreys' scale. }
    \label{fig:AIC_h0_s8}
\end{figure}

\subsection{Combined alleviation of the $H_0$ and $S_8$ tension}
\label{sec:H0_S8_tension}

The results for models' $\Delta$AIC that take both $H_0$ (R21) and $\sigma_8$ (V09) measurements into account are shown in Fig.~\ref{fig:AIC_h0_s8}. 
Based on the resulting $\Delta$AIC values, we find that the models that successfully alleviate the $H_0$ tension, discussed in Sec.~\ref{sec:H0_tension}, do not also alleviate the $S_8$ tension of V09 at the same time. 
We also tested the models' $\Delta$AIC values for other $S_8$ tensions, SZ21, DES, and SZ13 with our baseline datasets (P18+BAO16+SN) and their subsets. Again, none of those tension datasets had greater preference for any of the models than V09, as given by their $\Delta$AIC. 
In the case of P18+R21+V09, $w$CDM and a nonzero curvature are the preferred models with $\Delta$AIC values crossing the ``strong" threshold. 
When considering P18+BAO16+R21+V09, $w$CDM maintains the first position with $\Delta{\rm AIC} = -10.3$ (equivalent to 170:1 odds). EDE and PMF are next, followed by $\Lambda$CDM+$N_{\rm eff}$ and a nonzero curvature; however, their $\Delta$AIC values are below the threshold.
Finally, when testing against P18+BAO16+SN+R21+V09, no model has a $\Delta$AIC value above the ``strong" threshold of  -6.91 on the Jeffreys' scale or 30:1 odds, 
and the case of a non-minimal neutrino mass has a positive $\Delta$AIC value, i.e., it is certainly less preferred than $\Lambda$CDM.
We find 4 models with the lowest $\Delta$AIC value as our preferred models, which includes $w$CDM ($\Delta{\rm AIC} = -5.1$), EDE ($\Delta{\rm AIC} = -4.8$), $\Lambda$CDM+$N_{\rm eff}$ ($\Delta{\rm AIC} = -3.5$), and PMF ($\Delta{\rm AIC} = -2.2$).
Note that a nonzero curvature also has a small negative $\Delta$AIC; however, we do not select it as the preferred model as its $\Delta{\rm AIC}\sim 0$. 

As with the $S_8$ tension alone, we also ran model minimizations for R21 plus SZ21, DES, and SZ13. None of those combined $H_0$ plus $S_8$ tension datasets had greater preference for any of the models than R21+V09 given, as by their $\Delta$AIC.

\begin{figure*}[t]
\centering
\includegraphics[width=0.32\textwidth]{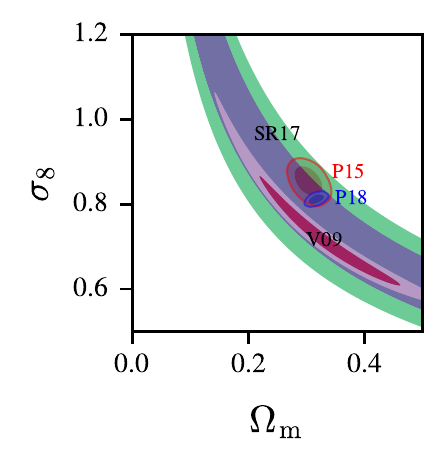}
\includegraphics[width=0.32\textwidth]{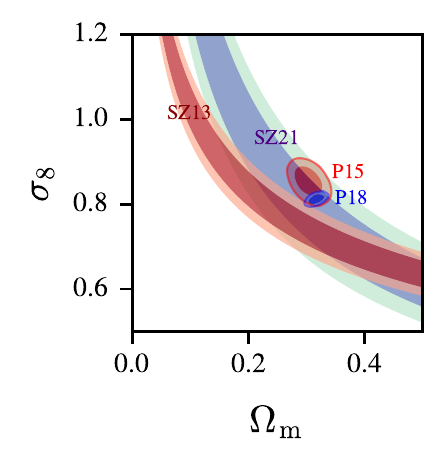}
\includegraphics[width=0.32\textwidth]{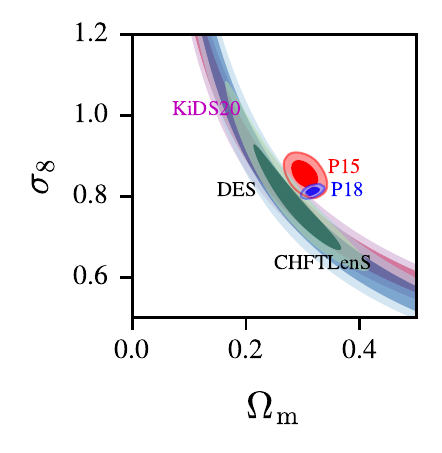}
\caption{Comparison of $1\sigma$ and $2\sigma$  contours of ($\Omega_\mathrm{m}$, $\sigma_8$) from Planck (2015 and 2018) and  measurements including X-ray clusters V09/SR17 (left panel), SZ clusters SZ13/SZ21 (middle panel) and lensing KiDS20/CHFTLenS/DES  (pink/olive/light blue, right panel).}
\label{fig:sigma8_tension}
\end{figure*}

\section{Discussion}
\label{sec:discussion}

\subsection{Cosmologies and Cosmological Parameters} 

\begin{figure*}[t]
    \centering
    \includegraphics[width=\textwidth]{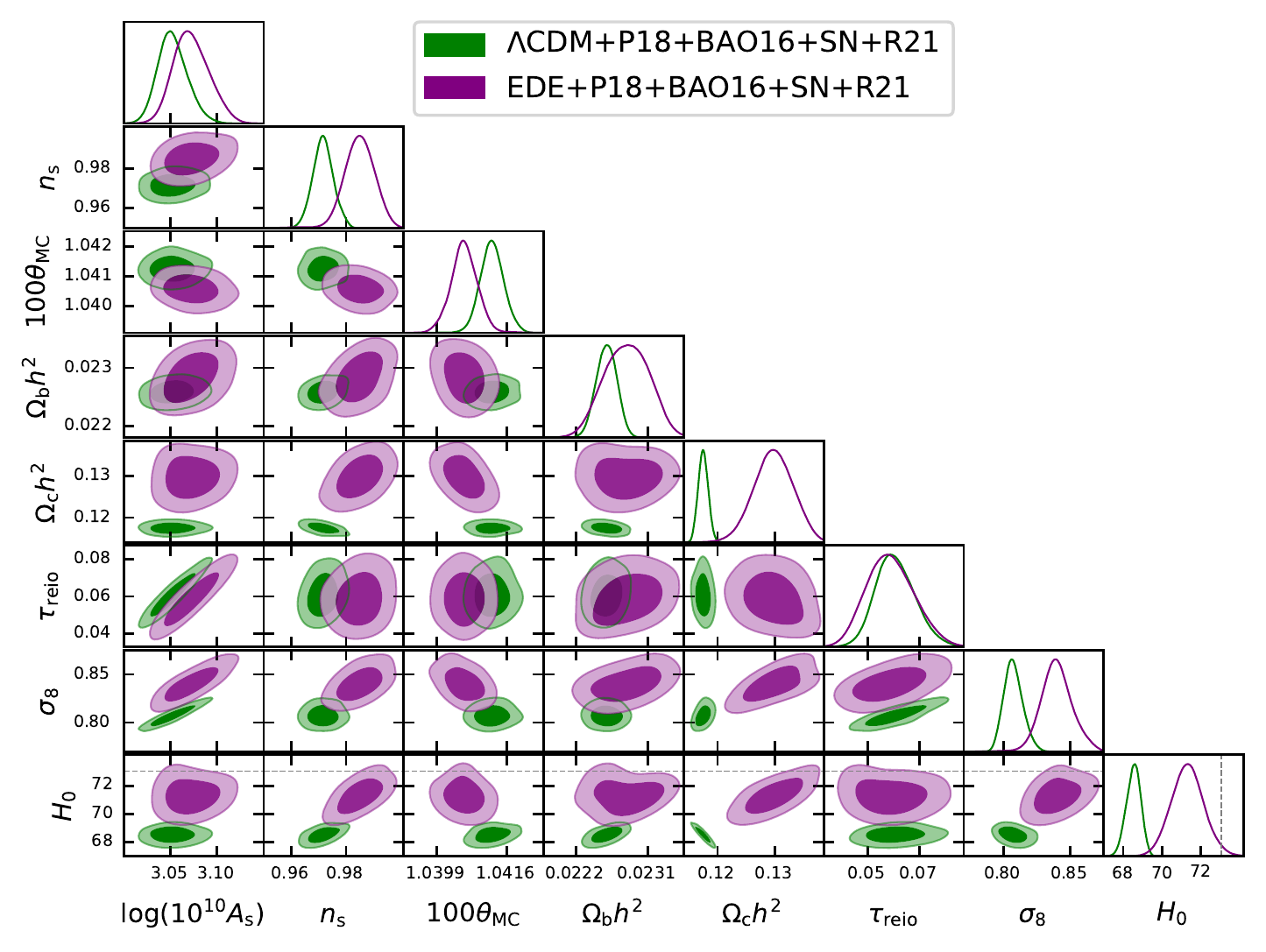}
    \caption{EDE v.s. $\Lambda$CDM: 68\% and 95\% C.L. contours of $\lbrace\log(10^{10}A_s)$, $n_s$, $100\theta_{\rm MC}$, $\Omega_b h^2$, $\Omega_c h^2$, $\tau_{\rm reio}$, $\sigma_8$, $H_0\rbrace$ of EDE (purple) v.s. $\Lambda$CDM (green) for our baseline cosmological datasets plus $H_0$, P18+BAO16+SN+R21. The gray dashed line denotes the best-fit value of $H_0 = 73.04\,{\rm km/s/Mpc}$ reported by R21.}
    \label{fig:tri_EDE}
\end{figure*}

\begin{figure*}[t]
    \centering
    \includegraphics[width=\textwidth]{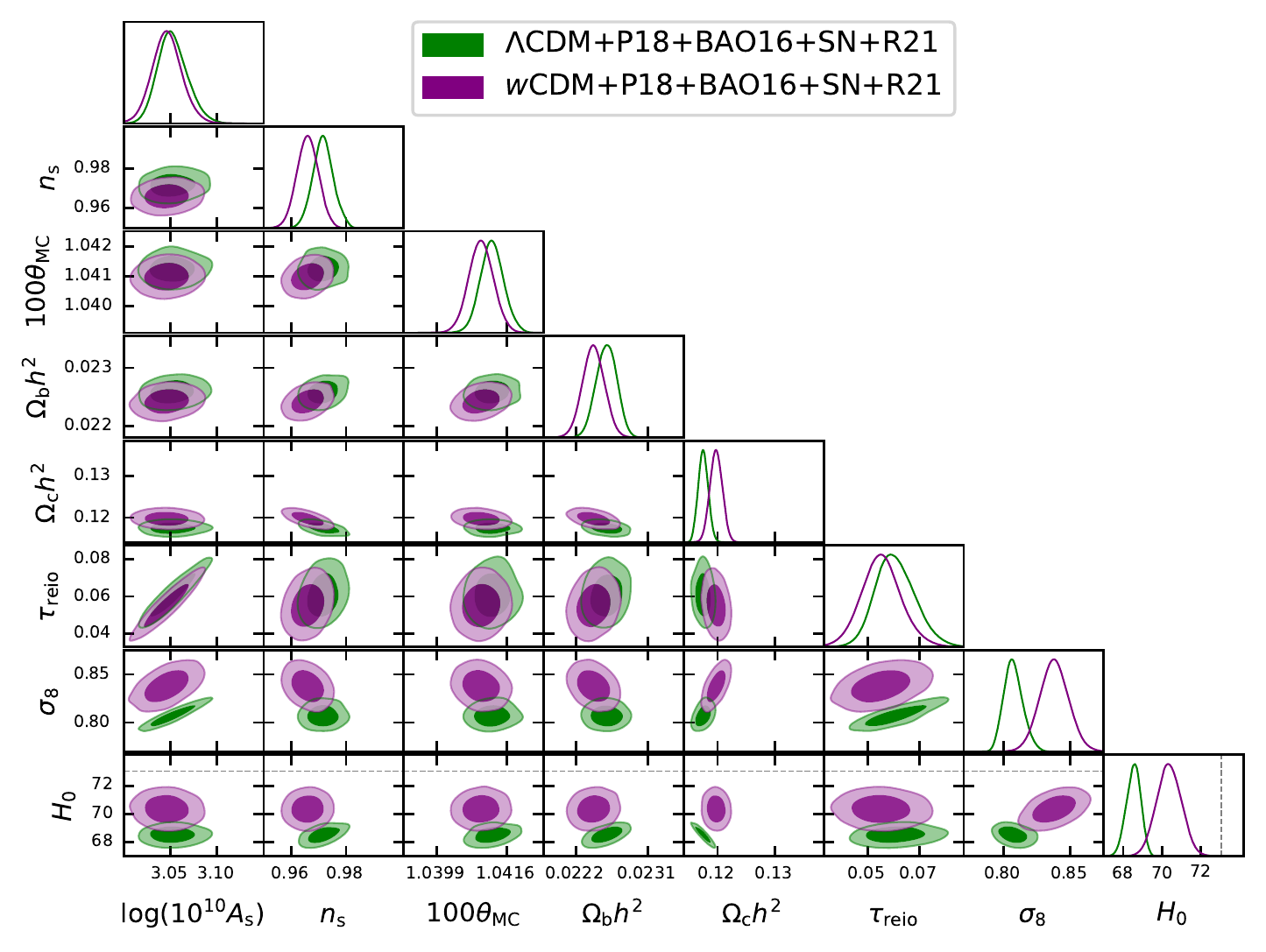}
    \caption{$w$CDM v.s. $\Lambda$CDM: 68\% and 95\% C.L. contours of $\lbrace\log(10^{10}A_s)$, $n_s$, $100\theta_{\rm MC}$, $\Omega_b h^2$, $\Omega_c h^2$, $\tau_{\rm reio}$, $\sigma_8$, $H_0\rbrace$ of $w$CDM  v.s. $\Lambda$CDM for our baseline cosmological datasets plus $H_0$, P18+BAO16+SN+R21. The gray dashed line denotes the best-fit value of $H_0 = 73.04\,{\rm km/s/Mpc}$ reported by R21.}
    \label{fig:tri_w}
\end{figure*}

\begin{figure*}[t]
\centering
\includegraphics[width=\textwidth]{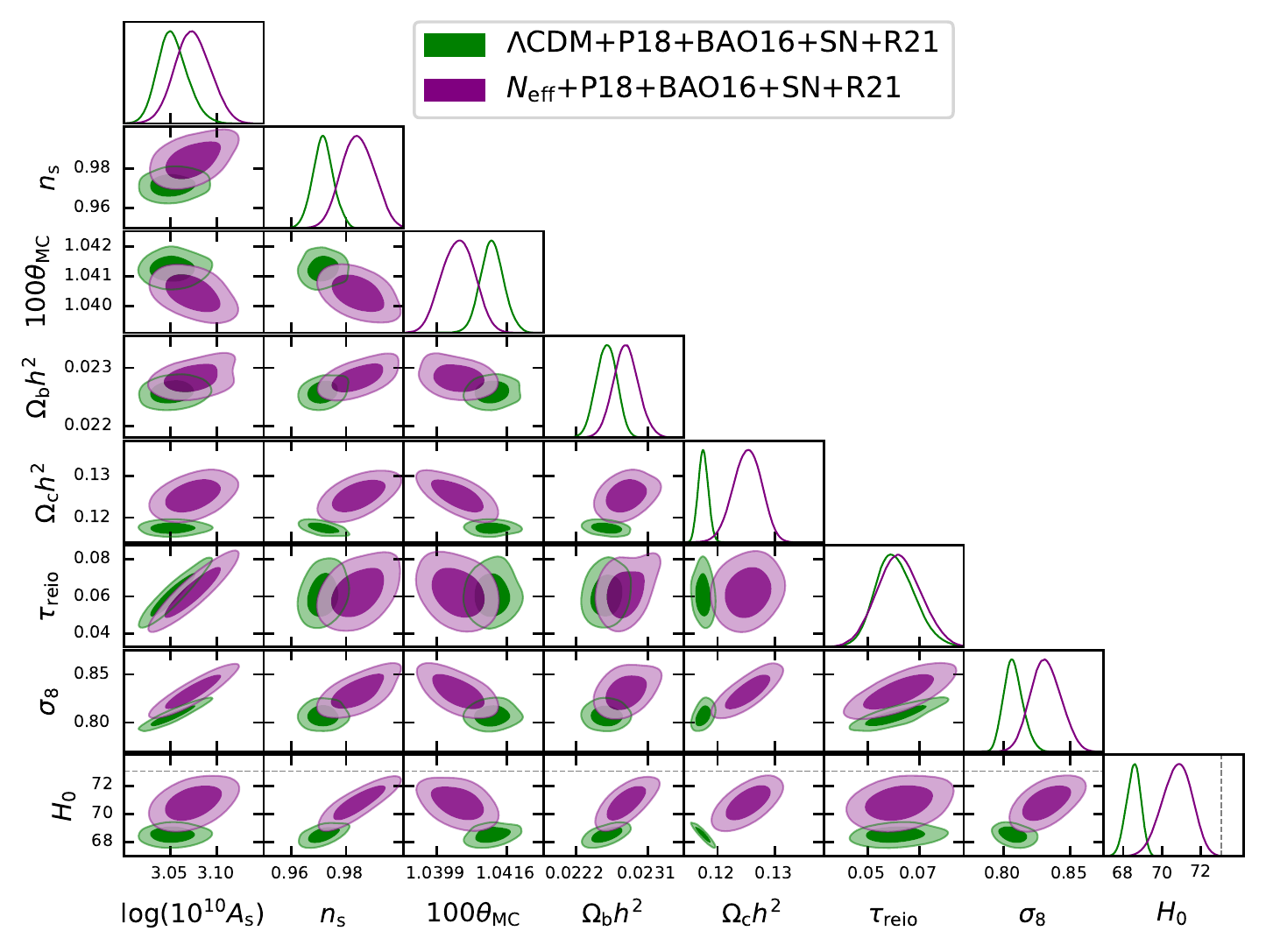}
\caption{$\Lambda$CDM+$N_{\rm eff}$ v.s. $\Lambda$CDM: 68\% and 95\% C.L. contours of $\lbrace\log(10^{10}A_s)$, $n_s$, $100\theta_{\rm MC}$, $\Omega_b h^2$, $\Omega_c h^2$, $\tau_{\rm reio}$, $\sigma_8$, $H_0\rbrace$ of $\Lambda$CDM+$N_{\rm eff}$ model (purple) v.s. $\Lambda$CDM (green) for our baseline cosmological datasets plus $H_0$, P18+BAO16+SN+R21.
The gray dashed line denotes the best-fit value of $H_0 = 73.04\,{\rm km/s/Mpc}$ reported by R21.
}
\label{fig:tri_Neff}
\end{figure*}

\begin{figure*}[t]
\centering
\includegraphics[width=\textwidth]{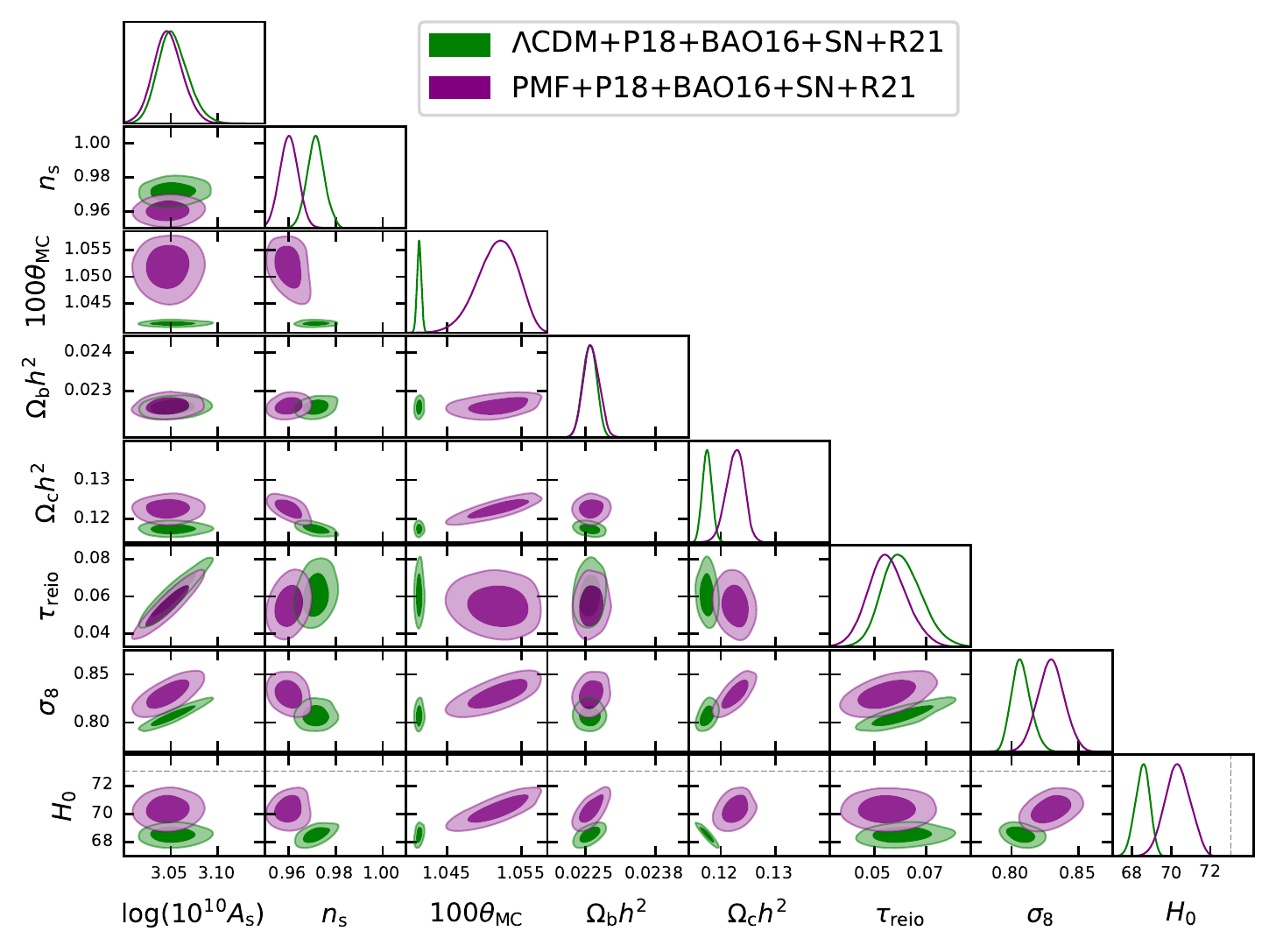}
\caption{PMF v.s. $\Lambda$CDM: 68\% and 95\% C.L. contours of $\lbrace\log(10^{10}A_s)$, $n_s$, $100\theta_{\rm MC}$, $\Omega_b h^2$, $\Omega_c h^2$, $\tau_{\rm reio}$, $\sigma_8$, $H_0\rbrace$ of PMF (purple) v.s. $\Lambda$CDM (green) for our baseline cosmological datasets plus $H_0$, P18+BAO16+SN+R21. The gray dashed line denotes the best-fit value of $H_0 = 73.04\,{\rm km/s/Mpc}$ reported by R21.}
\label{fig:tri_bar}
\end{figure*}

We further study the four preferred models selected in the previous section by running MCMC chains with the joint dataset P18+BAO16+SN+R21.
In Fig.~\ref{fig:tri_EDE}--Fig.~\ref{fig:tri_bar}, we present the contours and posterior distributions of cosmological parameters for each preferred model $\mathcal{M}$ with purple (green) contours corresponding to the result of $\mathcal{M}$ ($\Lambda$CDM).
The baseline cosmological parameters are the baryon density $\Omega_b h^2$, the CDM density $\Omega_c h^2$, the scalar spectral index $n_s$, the optical depth to reionization $\tau_{\rm reio}$, the \verb|CosmoMC| approximation to the angular size of the sound horizon $100\theta_{\rm MC}$, and the amplitude of primordial scalar perturbations $\log(10^{10} A_s)$. We also show the derived parameters of interest, $H_0$ and $\sigma_8$, and the summary of their mean values and $68\%$ C.L. intervals is given in Table~\ref{tab:LCDMcompare}.
For the $w$CDM model, the optical depth slightly decreases with respect to $\Lambda$CDM, as well as the scalar spectral index, the angular acoustic scale and the baryon density. The physical dark matter density increases noticeably, as well as the two derived parameters $\sigma_8$ and $H_0$. 

In the case of the $\Lambda$CDM+$N_{\rm eff}$ model, $\log(10^{10}A_S)$, the scalar spectral index, physical baryon density and physical dark matter density shift toward higher values than the ones of the standard $\Lambda$CDM. The values of $\sigma_8$ and $H_0$ are also greater. The optical depth to the epoch of reionization does not vary significantly. The angular acoustic scale is reduced with respect to $\Lambda$CDM.

For EDE we observe similar shifts as the one just described for the $N_{\rm eff}$ model. Moreover, we highlight that these two models have a higher $\sigma_8$ and higher $n_s$, which gives more early halo evolution. In Ref.~\cite{Klypin:2020tud}, they used a large suite of cosmological N-body simulations to explore the implications of the different cosmology implied by the EDE model of Ref.~\cite{Smith:2019ihp}. Given that the cosmological parameters are similar for EDE and $N_{\rm eff}$ cosmologies, the implications for structure formation may be similar. Namely, the increase in $\sigma_8$, $n_s$ and decrease in $\Omega_\mathrm{m}$, all enhance early galaxy formation, which may be indicated by the large number of massive galaxies being detected by JWST~\cite{2022arXiv220709436C,2022arXiv220712474F}. 

For the existence of PMF, and commensurate baryonic inhomogeneity in the early Universe, the shifts in some of the parameters are noticeable, in which the most significant one is the increased angular acoustic scale in \verb|CosmoMC| approximation $\theta_{\rm MC}$. However, we note that since the redshift of CMB photon decoupling is changed in PMF paradigm, $\theta_{\rm MC}$ is no longer a good approximation to the actual angular scale of the sound horizon $\theta_*$.
We check that the value of $\theta_*$ derived from the best-fit value of parameters in Table~\ref{tab:LCDMcompare} is consistent with P18.
The dark matter density is also higher in this model, as well as the two derived parameters $\sigma_8$ and $H_0$. The spectral index is reduced as well as the optical depth to the epoch of reionization. $\log(10^{10}A_s)$ and the baryon density do not suffer important changes from the standard values.

Even though the existence of a nonzero curvature is not part of our preferred models, we want to highlight that it is a successful model when only considering P18 datasets as well as the $H_0$ prior or $\sigma_8$ constraint. Nevertheless, when the additional datasets of BAO16 and SN are included, this model is no longer preferred due to the enhanced constraints on curvature from these data.

For SI$\nu$ (i.e., $\Lambda$CDM+$N_{\rm eff}$+$G_{\rm eff}$), we retrieve the result shown in previous literature that the likelihood peaks at moderate and strong interaction levels and moderate interaction levels are preferred by the baseline datasets.  We further show that moderate interaction level can effectively alleviate the $H_0$ tension when being tested against the baseline datasets with R21. However, as a result of an extra free parameter and no significant improvement in the fit, the AIC value of SI$\nu$ is in general poorer than the $\Lambda$CDM+$N_{\rm eff}$ case, thus SI$\nu$ is not selected as a preferred model.

\begin{table}[h]
\begin{tabular}{|l|c|c|c|}
\hline
Models            & \multicolumn{1}{l|}{$H_0$ (R21)}   & \multicolumn{1}{l|}{$\sigma_8$ (V09)} & \multicolumn{1}{l|}{$\sigma_8$ (DES)} \\ \hline
$\Lambda$CDM               & 4.63                         & 3.59                                  & 2.44                                  \\ \hline
EDE                & 1.83 & 4.21                                  & 2.93                                  \\ \hline
$w$CDM       & 3.16                         & 3.94                                  & 2.36                                  \\ \hline
$\Lambda$CDM+$N_{\rm eff}$ & 3.33                         & 3.61                                  & 2.43                                  \\ \hline
PMF                & 3.63                         & 4.01                                 & 2.90                                  \\ 
\hline
\end{tabular}
\caption{We show the $\sigma$ value of the residual tension of each model given our full baseline dataset P18, BAO16, and SN, when including the SHOES collaboration (R21) $H_0$ measurement, X-ray clusters measurement of $\sigma_8(\Omega_\mathrm{m}/0.25)^{0.47}$ in V09, and the DES measurement of $S_8$. Though we use more data from V09 and DES in our full analysis, here we use only the tension data point for the residual, which is calculated using Eq.~\eqref{eq:3}. \label{tab:sigmatension}} 
\end{table}

\begin{table}[h]
\begin{tabular}{|l|c|c|c|}
\hline
Models            & \multicolumn{1}{l|}{P18}   & \multicolumn{1}{l|}{P18+BAO16} & \multicolumn{1}{l|}{P18+BAO16+SN} \\ \hline
$\Lambda$CDM               & 0.8055   &    0.8062 &    0.8034                                 \\ \hline
EDE                & 0.8648   &0.8570 &  0.8391                     \\ \hline
$w$CDM       & 0.8164   &0.8235 	&0.8268                             \\ \hline
$\Lambda$CDM+$N_{\rm eff}$ & 0.8177   &0.8202 &0.8252                                  \\ \hline
$\Lambda$CDM+$\Sigma m_\nu$ & 0.8191   &0.8140 &0.8153                                \\ \hline
$\Lambda$CDM+$\Omega_k$ & 0.8164   &0.8193 &0.8168                                 \\ \hline
PMF                & 0.8231   &0.8147 &0.8253                                  \\\hline 
SI$\nu$               & 0.8189   &0.8301 &0.8254                                 \\ 
\hline
\end{tabular}
\caption{Best-fit values of the parameter $\sigma_8$ for the three combinations of our baseline datasets P18, P18+BAO16, and P18+BAO16+SN, when including the SHOES collaboration (R21) $H_0$ measurement for each of the analyzed models. The conclusions from the values $S_8$ are the same.\label{tab:sigma8value}}
\end{table}

\subsection{The Level of the $S_8$ Tension}
As described above, there is a variety of levels of tension in $S_8$ ($\sigma_8$) given by different datasets. Here we review the implications of adopting the various datasets, which provide a range of results, from no tension to appreciable tension, up to 4 $\sigma$. As we describe in this section, the most recent analyzes of X-ray, SZ, and optically selected clusters have no tension with the inferred amplitude of matter clustering from Planck 2018. Recent weak lensing datasets remain in tension with Planck 2018 at the level of $\sim\!2\sigma$. In order to understand the nature of the $S_8$ ($\sigma_8$) tension further, we show $\sigma_8$ v.s. $\Omega_\mathrm{m}$ contours in Fig.~\ref{fig:sigma8_tension}, comparing Planck 2015 (P15)~\cite{Planck:2015fie} and P18 with datasets that constrain $S_8$ including X-ray clusters (left panel),  Sunyaev-Zel'dovich (SZ) clusters (middle panel) and weak lensing measurements (right panel). 
For X-ray clusters, we show V09~\cite{Vikhlinin:2008ym} introduced in Sec.~\ref{sec:dataset} and HIFLUGCS (SR17)~\cite{Schellenberger:2017wdw} which is a cluster sample measurement of the brightest 64 X-ray galaxy clusters that has individually determined, robust total mass estimates, and compares with Planck-SZ determined mass estimates. Their ensuing $S_8$ constraint is  $\sigma_8 \sqrt{\Omega_\mathrm{m} /0.3}  = 0.792\pm 0.049$. As can be seen in this figure, the tension is relaxed both by P18 shifting lower than P15 and by the updated X-ray cluster constraints shifting higher.\footnote{P18 shifts in this parameter space with respect to P15 due to the change in the determination of the optical depth given the updated CMB polarization measurements of P18~\cite{Planck:2018vyg}.}
For SZ clusters, we show the constraint derived from the sample of 189 galaxy clusters from the Planck SZ catalog (SZ13)~\cite{Planck:2013lkt}.
We compare SZ13 with recent results from the SPT-SZ collaboration (SZ21)~\cite{SPT:2021efh}. Similar to X-ray clusters, the tension in SZ cluster samples is relaxed both by P18 shifting lower than P15 and by the newer SZ cluster constraints shifting higher in this parameter space. 
Optically selected cluster samples also determine a higher value for $\sigma_8$,  consistent with noncluster probes \cite{Abdullah:2020qmm}.

For lensing measurements, we compare results from the tomographic weak lensing analysis of the Canada-France-Hawaii Telescope Lensing Survey (CFHTLenS)~\cite{Heymans:2013fya}, cosmic shear analysis of the fourth data release of the Kilo-Degree Survey (KiDS-1000)~\cite{KiDS:2020suj} and the most recent results from the combined galaxy clustering and lensing measurement of DES Year 3~\cite{DES:2021bwg}. 
As discussed in Sec.~\ref{sec:dataset}, DES has robust photometric redshift calibration methods able to recover results obtained from simulated surveys. Due to the size of the dataset, the errors on the cosmological parameters are also reduced relative to previous datasets. 

The tension from V09 is clear, with an approximately $3.59\sigma$ deviation with Planck, cf. Table~\ref{tab:sigmatension}. %
Recall we adopted this as a benchmark dataset in order to introduce high-tension and therefore potentially infer the most likely $\Lambda CDM$ model extension.
However, updated measurements such as SR17 and SZ21 yield a larger value of $\sigma_8$ which is consistent with P18.
On the other hand, lensing measurements largely agree with each other. 

The lack of a preference for a nonzero neutrino mass when including low-$S_8$/$\sigma_8$ datasets with P18 is due to the shift in the CMB optical depth and scalar amplitude parameters to be significantly lower with P18's updated polarization anisotropy measurements, relative to earlier Planck data, along with slight shifts in the other parameters. Overall, as discussed in Sec.~\ref{sec:result}, the $S_8$ tension is no longer indicative of a possible non-minimal neutrino mass. As we have seen in the evolution of the X-ray and SZ cluster data, the tension with those datasets has been alleviated. This leaves the weak-lensing based inferences of $S_8$. For the case of DES, the tension with $\Lambda$CDM is mild, at $\sim\!2.44\sigma$ (see Table~\ref{tab:sigmatension}).

Most importantly, we find that none of the new models considered alleviate the $S_8/\sigma_8$ tension at the same time as the $H_0$ tension (c.f. Table~\ref{tab:sigma8value}). When including R21, the value of $S_8/\sigma_8$ is reduced even in the case of $\Lambda$CDM. All of the models considered drive $\sigma_8$ higher, as shown in Table~\ref{tab:sigma8value}, a feature prevalent in many models trying to alleviate both tensions \cite{Jedamzik:2020zmd}. The values for $S_8$ are also higher for all models, except for the $\Lambda$CDM+$\Omega_k$ model, and only for the case of considering P18 data plus R21 alone. None of these new models alleviate the $S_8/\sigma_8$ tension better than $\Lambda$CDM when including our full dataset. Therefore, when considering if a new model does better than $\Lambda$CDM in alleviating the $S_8$ problem simultaneously with the $H_0$ problem, one should compare to $\Lambda$CDM+R21's own alleviation of $S_8$, and not $\Lambda$CDM without $H_0$ information.

\section{Conclusion}
\label{sec:conclusion}
In this paper, we have investigated how well seven models---$w$CDM, $\Lambda$CDM+$N_{\rm eff}$, $\Lambda$CDM+$\sum m_\nu$, $\Lambda$CDM+$\Omega_k$, PMF, SI$\nu$ and EDE---explain or fail to explain the $H_0$ and $S_8$ tensions. We do this by calculating both the change in the AIC and the change in the total $\chi^2$ for the models and datasets.  We find that EDE, $w$CDM, PMF, and $\Lambda$CDM+$N_{\rm eff}$ pass the threshold of the ``strong'' preference criterion of $\Delta\mathrm{AIC}<-6.91$.  However, each of these models still has a residual tension with the R21 $H_0$ constraint of greater than $3\sigma$ except for EDE, which has a residual tension of less than $2\sigma$. Inclusion of more model parameters corresponding to greater details of the SI$\nu$, EDE and PMF models could lead to a poorer indication for their preference by $\Delta$AIC, but an exploration of those extensions is beyond the scope of this work~\cite{Rashkovetskyi:2021rwg,Thiele:2021okz}. 
  Therefore, of the seven models, EDE satisfies both in having an overall better fit to all the data, including $H_0$ as well as having almost no remaining tension with the single measurement of $H_0$. The better fit of EDE largely comes from its consistency with periodic features the high-$\ell$ $C_\ell$ measurements of Planck 2018~\cite{Smith:2019ihp}, which will probed well by upcoming CMB experiments~\cite{SPT-3G:2021vps,Thornton:2016wjq,CMB-S4:2016ple}.

For the case of the $S_8/\sigma_8$ tension, we adopted a strong-tension dataset (V09), but our conclusions do not change with other $S_8$ datasets. Only in the case of Planck 2018 CMB data plus V09, are evolving dark energy $w>-1$ and $\Omega_k$ not disfavored relative to $\Lambda$CDM. However, with the BAO16+SN data, no model alleviates the $S_8$ tension, due to those constraints on the expansion history. We discussed how the $S_8$ tension has been alleviated to the $\sim\!2\sigma$ level both by shifts in the Planck 2015 to 2018 analyzes, as well as shifts in structure formation measures of $S_8/\sigma_8$, whether by x-ray clusters, SZ clusters, or weak lensing. Importantly, we show that a nontrivial neutrino mass ($\Sigma m_\nu>0.06\,\mathrm{eV}$), does not alleviate the $S_8/\sigma_8$ tension. 

Significantly, we showed that adding the $H_0$ measurement of R21 to all of the datasets we considered substantially lowers $S_8/\sigma_8$ for $\Lambda$CDM due to a shift to a larger $\Omega_\Lambda$, and therefore a commensurate suppression in the growth of the large scale structure. Importantly, \textit{no model considered here} lowers the best-fit value of $S_8/\sigma_8$ better than $\Lambda$CDM when including the $H_0$ (R21) tension. Therefore, claims in other work of models alleviating both $H_0$ and $S_8$ tensions should be sure to compare with $\Lambda$CDM's own $S_8$ alleviation when including $H_0$, and not compare to $\Lambda$CDM without the $H_0$ constraint.

In the context of Bayesian model selection, via the AIC, the observation of the $H_0$ tension updates our belief such that the EDE model is best.
Given the fact that the values of the $S_8/\sigma_8$ parameters become higher for all models when the R21 observation is included, the observation of the $S_8/\sigma_8$ tension does not update the data's preference for the EDE model.  For the kinds of models still allowed by the joint P18+BAO16+SN+R21 datasets, the $H_0$ and $S_8/\sigma_8$ tensions will pull the models in different directions.

Future tests of an EDE epoch could come from high-$\ell$ CMB  measurements, as discussed earlier, or from the turn over in the matter power spectrum at very large scales, which should constrain $\theta_{\rm eq}$, the angular size of the sound horizon at matter-radiation equality that could be constrained by large-scale structure surveys. And, maybe most importantly, future independent determinations of the local expansion history $H_0$ may reaffirm its tension or relax it.

\begin{table*}[t]
\begin{center}
\begingroup
\openup 8pt
\newdimen\tblskip \tblskip=10pt
\nointerlineskip
\vskip -4mm
\small
\setbox\tablebox=\vbox{
    \newdimen\digitwidth
    \setbox0=\hbox{\rm 0}
    \digitwidth=\wd0
    \catcode`"=\active
    \def"{\kern\digitwidth}
    \newdimen\signwidth
    \setbox0=\hbox{+}
    \signwidth=\wd0
    \catcode`!=\active
    \def!{\kern\signwidth}
\halign{
\hbox to 1.1in{$#$\leaderfil}\tabskip=2em&$#$\hfil&$#$\hfil&$#$\hfil&$#$\hfil&$#$\hfil&\hfil$#$\hfil\tabskip=0pt\cr
\noalign{\doubleline}
\multispan1 ~~~~~~~Models \hfil&\multispan1\hfil $w$CDM \hfil&\multispan1\hfil EDE \hfil&\multispan1\hfil $\Lambda$CDM+$N_{\rm eff}$\hfil&\multispan1\hfil PMF \hfil\cr
\noalign{\vskip -2pt}
\omit Parameters\hfil&\omit\hfil 68\% limits\hfil&\omit\hfil 68\% limits\hfil&\omit\hfil 68\% limits\hfil&\omit\hfil 68\% limits\hfil\cr
\noalign{\vskip 3pt\hrule\vskip 5pt}
\Omega_{b} h^2&0.02242 \pm 0.00014& 0.02284 \pm 0.00027 &0.02282^{+0.00013}_{-0.00014}&0.02261 \pm 0.00015\cr
\Omega_{c} h^2&0.1197\pm0.0011 &0.1298 \pm 0.0034 &  0.1252 \pm 0.0023& 0.1226 ^{+0.0017}_{-0.0015}\cr
100\theta_{\mathrm{MC}}&1.04098 \pm 0.00030  & 1.04057\pm0.00033 &  1.04042 \pm 0.00037   &  1.0518 ^{+0.0030}_{-0.0024}\cr
\tau&0.0553 \pm 0.0078 & 0.0587^{+0.084} _{-0.010} & 0.0621 \pm 0.0081  &  0.0550^{+0.0069}_{-0.0078}\cr
\ln(10^{10} A_s)&3.046 \pm 0.016&3.072 ^{+0.017} _{-0.021}  &  3.074 \pm 0.018&3.074 \pm 0.015 \cr
n_s &0.9660 \pm 0.0039 & 0.9848 \pm 0.0056  & 0.9843 \pm 0.0057 &  0.9602 \pm 0.0039     \cr
H_0\,[{\rm km}\,{\rm s}^{-1}\,{\rm Mpc}^{-1}]&70.36 \pm 0.64  & 71.31\pm0.83 &  70.77^{+0.82}_{-0.72} &  70.33 \pm 0.64\cr
\Omega_{m}& 0.2886 \pm 0.0057& 0.3014 \pm 0.0062& 0.2969\pm0.0048& 0.2950\pm0.0052\cr
\sigma_8&0.838 \pm 0.011  & 0.840^{+0.011} _{-0.012}  &  0.831 \pm 0.011    & 0.8294 \pm 0.0098 \cr
\text{extra param.}& w = -1.094 \pm 0.026  & f_{z_c}=0.096^{+0.023}_{-0.027} &  N_{\rm eff}=3.48 \pm 0.12  & b = 0.57 \pm 0.19 \cr
&    & z_{c}=3090 \pm 38 &    &  \cr
&    &  \Theta_i = 1.9^{+1.0}_{-1.8}&    &   \cr
\noalign{\vskip 5pt\hrule\vskip 3pt}
} 
} 
\endPlancktable
\endgroup
\end{center}
\caption{Best-fit values and $68\%$ intervals of parameters for the preferred candidate models from
P18+BAO16+SN+R21. The first six rows are the baseline parameters of $\Lambda$CDM, which are sampled in the MCMC analysis with flat priors. The next three rows are the derived parameters $H_0$, $\Omega_\mathrm{m}$ and $\sigma_8$. The last row represents model parameters in addition to $\Lambda$CDM.
}
\label{tab:LCDMcompare}
\end{table*}

\section*{Acknowledgment}

We thank Levon Pogosian for sharing the \verb|CAMB| code implemented with baryon inhomogeneity and for useful discussions. We acknowledge useful comments on the manuscript from Karsten Jedamzik, Anatoly Klypin, Levon Pogosian, Nils Sch\"oneberg, and Radoslaw Wojtak. K.N.A. acknowledges useful discussions with John Carlstrom, Wayne Hu, and Joel Primack. J.L.K. and K.N.A. are supported by U.S. National Science Foundation (NSF) Theoretical Physics Program, Grants No. PHY-1915005 and No. PHY-2210283.

\appendix

\section{Details of considered models}
\label{app:model}

\subsection{Early dark energy}

The phenomenology of EDE can be realized by, e.g., a scalar field $\phi$ with a potential $V(\phi) \propto \left[ 1- \cos(\phi/f)\right]^n$ such that $\phi$ is frozen at $z < z_c$ with $w_\phi = -1$ and starts to oscillate at $z = z_c$ and can be effectively described as a fluid with $w_\phi = w_n = (n-1)/(n+1)$, or, slow-roll of $\phi$ down a potential that $V(\phi) \propto \phi$ at $z< z_c$ and $V(\phi)\rightarrow 0$ at late time; see~\cite{Kamionkowski:2014zda,Karwal:2016vyq,Poulin:2018dzj,Poulin:2018cxd,Smith:2019ihp} for further details.
Note that whether EDE works does not depend much on the details of potential, as long as the typical evolution of energy density is fulfilled. 

Taking the scalar potential $V(\phi) \propto \left[ 1- \cos(\phi/f)\right]^n$ as the benchmark model, in the fluid approximation the evolution of density parameter of $\phi$ reads~\cite{Poulin:2018dzj}
\begin{align}
\Omega_\phi(a) = \dfrac{2\Omega_\phi (a_c)}{(a/a_c)^{3(w_n + 1)}+1}\,,
\end{align}
where $a$ is the scale factor of the Universe and $a_c = (1+z_c)^{-1}$.
The equation of state is 
\begin{align}
  w_\phi (a) = \dfrac{1+w_n }{1+(a_c/a)^{3(1+w_n)}} -1\,,
\end{align}
such that the energy density of $\phi$ can dilutes faster than radiation at $z < z_c$ for $n \geq 3$; in this work, we focus on the case that $n =3$.
Following the literature~\cite{Smith:2019ihp}, we define $f_{z_c} \equiv \Omega_\phi (a_c)/\Omega_{\rm tot} (a_c)$ with $\Omega_{\rm tot} (a_c)$ being the density parameter of the total energy density at $z = z_c$ and $\Theta \equiv \phi/f$ being the renormalized field variable which determines the effective sound speed. In our analysis, we set $z_c$, $f_{z_c}$ and the initial value of the renormalized field variable $\Theta_i$ as free parameters.

\subsection{Self-interacting neutrinos}

We focus on the particle model where the self-interaction of neutrinos is mediated by a massive scalar $\phi$ with mass $m_\phi$; the coupling strength of $\phi\nu\nu$ is $g_\nu$.
When the neutrino temperature $T_\nu \ll m_\phi$, the scalar particle can be integrated out and the interaction can be described by the effective field theory (EFT).
In the EFT framework, the self-interaction $\nu \nu \rightarrow \nu\nu$ is analogous to the 4-Fermi interaction with a constant $G_{\rm eff} \equiv g_\nu^2 /m_\phi^2$. \footnote{Note that $G_{\rm eff}$ defined here is equivalent to $ G_\nu$ defined in Refs.~\cite{Cyr-Racine:2013jua,Lancaster:2017ksf}.}
In the limit that neutrinos are relativistic, i.e., $T_\nu \gg m_\nu$, the thermal-averaged cross section of self-interaction reads
\begin{align}
\langle \sigma v \rangle  \sim G_{\rm eff}^2 T_\nu^2\,.
\end{align}
The interaction rate can be written as $\Gamma = n_\nu \langle \sigma v \rangle \sim  G_{\rm eff}^2 T_\nu^5$ as $n_\nu \propto T_\nu^3$.
We focus on the strongly interacting regime  $G_{\rm eff} \gg G_F \simeq 1.2\times 10^{-5}\,{\rm GeV}^{-2}$ with $G_F$ being the Fermi constant, such that the self-interaction can drastically delay the SM neutrino decoupling and affect the CMB angular power spectra.

To consistently solve for the CMB angular power spectra in the SI$\nu$ scenario, when evolving the perturbations, we need to augment the usual Boltzmann hierarchy of neutrinos in the cosmological perturbation theory with additional damping terms. 
The phase-space distribution function of neutrinos reads $f= f_0 (1+\Psi)$ with $f_0$ being the unperturbed one. 
The perturbation $\Psi$ can be expanded into a Legendre series; for each multipole $\ell$, the corresponding amplitude is $\Psi_\ell$. 
Therefore, instead of solving the Boltzmann equation in terms of $\Psi$, we can evolve the Boltzmann hierarchy in $\Psi_\ell$.
Following the convention in~\citet{Ma:1995ey}, in the synchronous gauge the Boltzmann hierarchy of massive neutrino can be written as 
\begin{align}
 \dot{\Psi}_0 &= - \dfrac{qk}{\epsilon} \Psi_1 + \dfrac{\dot{h}}{6} \dfrac{d\ln f_0}{d\ln q}\,, \\
 \dot{\Psi}_1 & = \dfrac{qk}{3\epsilon}{\Psi_0  - 2\Psi_2}\,, \\
 \dot{\Psi}_2 &= \dfrac{qk}{5\epsilon} (2\Psi_1 -3 \Psi_3) -\left( \dfrac{\dot{h}}{15}+\dfrac{2\dot{\eta}}{5}\right) \dfrac{d\ln f_0}{d\ln q} + C_2^{\rm damp}\,,\\
 \dot{\Psi}_{\ell\geq 3} &= \dfrac{qk}{(2\ell+1)\epsilon} (\ell\Psi_{\ell-1} - (\ell+1) \Psi_{\ell+1}) + C_{\ell\geq 3}^{\rm damp}\,,
\end{align}
where overdot stands for derivative with respect to the conformal time, $k$ is the comoving wave number, $q$ and $\epsilon$ are the comoving momentum and energy, $h$ and $\eta$ are fields describing the metric perturbation.
Note that collision terms stemmed from the self-interaction does not affect $\ell= 0$ and $\ell=1$ since number density and energy density are conserved.

The complete formulas of damping terms $C_\ell^{\rm damp}$ derived from the integral of collision terms can be found in~\cite{Oldengott:2014qra,Oldengott:2017fhy}. 
In this work we adopt the relaxation time approximation~\cite{PhysRev.94.511,Hannestad:2000gt} or the separable ansatz~\cite{Cyr-Racine:2013jua} which gives 
\begin{align}
C^{\rm damp}_\ell = \alpha_\ell \dot{\tau_\nu} \Psi_\ell\,,
\end{align}
where $\alpha_\ell$ is the numerical factor from the integration over momentum and $\dot{\tau_\nu} = -a\Gamma$ is the rate of change of the neutrino opacity with $a$ being the cosmological scale factor.
This approximation is shown to be an adequate description of a system without dissipative process~\cite{Bazow:2016oky} and numerically agrees with the full treatment of self-interaction if the correct $\alpha_\ell$ is taken for each $\ell$~\cite{Oldengott:2017fhy}.\footnote{To avoid confusion, we note that in~\cite{Cyr-Racine:2013jua,Lancaster:2017ksf}, $G_{\rm eff} \equiv \sqrt{\alpha_\ell} G_\nu  =\sqrt{\alpha_\ell} g_\nu^2/m_\phi^2$.}

We modify \verb|CAMB| to include the effect of SI$\nu$; see~\cite{Das:2020xke} for existing codes.
At the early time, we approximate neutrinos as a perfect fluid to avoid a numerically stiff problem.
A perfect fluid has no stress; therefore, if $\Gamma \gg H$ is met, we set $\Psi_{\ell\geq 2} = 0$ in the initial condition and only evolve $\Psi_{\ell= 0,1}$.
As the Universe expands, eventually $\Gamma$ drops below $H$ and we start to solve for the full Boltzmann hierarchy once it is numerically solvable.

The nonstandard interaction of SM neutrinos faces constraints in aspects of particle model, cosmology, astrophysics and laboratory experiments. 
Theoretically, the validity of perturbative calculations requires $g_\nu \leq 4\pi$.
The thermal population of neutrinos and $\phi$ can be probed by $N_{\rm eff}$ at big bang nucleosynthesis (BBN) and CMB epochs, resulting in constraints on the underlying particle nature; especially, for the case where $\phi$ is a real scalar $m_\phi > 1.3 \,{\rm MeV}$ and the Dirac nature of neutrinos is strongly constrained~\cite{Blinov:2019gcj}.
Modification of the neutrino free-streaming leads to deviations from the standard cosmology in the CMB angular power spectrum, giving us a leverage to constrain the self-interaction strength~\cite{Archidiacono:2013dua,Cyr-Racine:2013jua,Lancaster:2017ksf,RoyChoudhury:2020dmd,Brinckmann:2020bcn}.
Furthermore, the propagation of energetic neutrinos is affected by the self-interaction; for example, detection of ultra-high energy neutrinos from supernovae~\cite{Manohar:1987ec,DICUS198984,Kolb:1987qy} or as cosmic ray~\cite{Keranen:1997gz,Hooper:2007jr,Ng:2014pca,Ioka:2014kca,Cherry:2014xra} can place a limit on $G_{\rm eff}$.
Finally, upon imposing a UV model for the effective interaction, strong bounds can also arise from SM precision observables, e.g., the $T$-parameter and decay of SM particles~\cite{Laha:2013xua,Blinov:2019gcj,Lyu:2020lps}.
In this work, we remain agnostic about the UV-origin of such effective interactions and consider the flavor-universal case for simplicity, which leaves us the only free parameter $G_{\rm eff}$. 
Recasting the result of $G_{\rm eff}$ into that of the $(g_\nu, m_\phi)$ parameter space is straightforward as long as proper UV models are applied.

\subsection{Primordial magnetic fields \& baryon inhomogeneity}

By enhancing the hydrogen recombination rate, baryon inhomogeneity before recombination can change the process of CMB photon decoupling. The degree of inhomogeneity is parameterized by the clumping factor
\begin{align}
b \equiv \dfrac{\langle n_b^2 \rangle  }{\langle n_b \rangle^2} - 1\,,
\end{align}
where $n_b$ is the baryon number density.
The baryon inhomogeneity can be due to, e.g., the existence of primordial magnetic fields~\cite{Jedamzik:2018itu}.

The generation of primordial magnetic fields can happen in the early Universe such as during phase transitions and during inflation; see~\cite{Durrer:2013pga} for detailed review.
By solving the evolution of the primordial magnetic field $\vec{B}$~\cite{Jedamzik:2013gua}, Ref.~\cite{Jedamzik:2020krr} found that $|\vec{B}| \sim \mathcal{O}(0.1\,{\rm nG})$ at Mpc scale to alleviate the Hubble tension by sourcing a baryon inhomogeneity $b \sim \mathcal{O}(0.1)$ that makes recombination happen earlier and thus reduces the sound horizon. In addition, such primordial magnetic fields can be the origin of the cluster magnetic fields observed today~\cite{Dolag:1999wvi,Banerjee:2003xk,Banerjee:2004df}.

We utilize a modified version of \verb|CAMB|, which includes the computation of the ionization fraction in three zones to account for a nonzero $b$, to evaluate the significance of baryon inhomogeneity in solving both the $H_0$ and $\sigma_8$ tensions.
We adopt M1 in~\cite{Jedamzik:2020krr} as a benchmark model for the three-zone calculation; see the mentioned reference for details of methodology and relevant parameters.

\bibliography{refs}
\end{document}